\documentclass[fleqn,usenatbib]{mnras}

\usepackage{newtxtext,newtxmath}

\usepackage[T1]{fontenc}
\usepackage{ae,aecompl}
\usepackage{threeparttable}

\usepackage{graphicx}	
\usepackage{amsmath}	
\usepackage{xcolor}

\title[Solo: The Local Group]{Solo dwarfs II: The stellar structure of isolated Local Group dwarf galaxies} 

\author[Higgs et al.]{C.R. Higgs$^{1}$\thanks{E-mail: higgs@uvic.ca}, A.W. McConnachie$^{2}$, N. Annau$^{1}$, M. Irwin$^{3}$, G. Battaglia$^{4,5}$,
P. C{\^o}t{\'e}$^{2}$, 
\newauthor G.F. Lewis$^{6}$, K. Venn$^{1}$\\
$^{1}$Physics \& Astronomy Department, University of Victoria, 3800 Finnerty Rd, Victoria, B.C., Canada, V8P 5C2\\
$^{2}$NRC Herzberg Astronomy and Astrophysics, 5071 West Saanich
  Road, Victoria, B.C., Canada, V9E 2E7\\
${^3}$Institute of Astronomy,  Madingley  Road,  Cambridge, CB3  0HA,  U.K.\\
${^4}$Instituto de Astrofísica de Canarias, Calle Vía Láctea s/n, E-38206 La Laguna, Tenerife, Spain \\
${^5}$Universidad de La Laguna. Avda. Astrofísico Fco. Sánchez, La Laguna, Tenerife, Spain\\
${^6}$Sydney Institute for Astronomy, School of Physics, A28, The University of Sydney, Sydney, NSW 2006, Australia\\
}

\date{Accepted XXX. Received YYY; in original form ZZZ}

\pubyear{2020}

\begin{document}
\label{firstpage}
\pagerange{\pageref{firstpage}--\pageref{lastpage}}

\maketitle

\begin{abstract}
The {\it Solo} ({\it So}litary {\it Lo}cal) Dwarf Galaxy survey is a volume limited, wide-field $g-$ and $i-$band survey of all known nearby (\textless 3\,Mpc) and isolated (\textgreater 300\,kpc from the Milky Way or M31) dwarf galaxies. This set of 44 dwarfs are homogeneously analysed for quantitative comparisons to the satellite dwarf populations of the Milky Way and M31. In this paper, an analysis of the 12 closest {\it Solo} dwarf galaxies accessible from the northern hemisphere is presented, including derivation of their distances, spatial distributions, morphology, and extended structures, including their inner integrated light properties and their outer resolved star distributions. All 12 galaxies are found to be reasonably well described by two-dimensional S\'{e}rsic functions, although UGC 4879 in particular shows tentative evidence of two distinct components.  No prominent extended stellar substructures, that could be signs of either faint satellites or recent mergers, are identified in the outer regions of any of the systems examined.

\end{abstract}

\begin{keywords}
galaxies: dwarf -- galaxies: general -- Local Group -- galaxies: photometry -- galaxies: structure -- galaxies: stellar content
\end{keywords}


\section{Introduction}

Dwarf galaxies are a fascinating and challenging regime in which to study galaxy evolution. Both internal and external processes dramatically impact the morphology and evolution of dwarfs as a consequence of their low masses and shallow gravitational potential wells. Star formation and the resulting feedback (e.g. \citealt{ElBadry2018}), ram pressure stripping (e.g. \citealt{Gunn1972, Grebel2003} 
), tidal stripping (e.g. \citealt{Fattahi2018}), tidal stirring (e.g. \citealt{Kazantzidis2011, Mayer2000}), reionization (e.g. \citealt{Ordonez2015, Wheeler2019}), mergers and interactions (e.g. \citealt{Deason2014}) are all processes which can significantly alter a dwarf's structure, morphology and stellar populations. The observational challenge then becomes understanding the complex interplay between initial formation processes and the internal and external processes that shape and reshape dwarfs. 

Dwarf galaxies are the most numerous class of galaxy in the Universe and the population of known nearby dwarfs has increased dramatically in recent years (e.g. numerous dwarfs discovered in the Dark Energy Survey \citealt{Bechtol2015,DrlicaWagner2015, Crnojevic2018, DrlicaWagner2020}; Crater III  and Bootes IV from the Hyper Suprime-Cam Subaru Strategic Program \citealt{Homma2018,Homma2019};  Sagittarius II, Draco II and Laevens 3 from the Pan-STARRS 1 $3\pi$ Survey \citealt{Laevens2015} among others). The parameter space that dwarfs are known to inhabit has been broadened by these discoveries and others (e.g. Crater 2 and Antlia 2  \citealt{Torrealba2016,Torrealba2019}). New terminology such as ultra-faint and ultra-diffuse is now in common usage \citep{Simon2019}. Many dwarf galaxies are so faint that the only detailed information obtainable for them comes through their resolved stars, with the result that the dwarfs in the nearby Universe are a critical observational sample for astronomers. 
 
Within the Local Group, the satellite dwarfs of the Milky Way (e.g., the MegaCam Survey of Outer Halo Satellites, \citealt{Marchi-Lasch2019}) and those around M31 (e.g., the Pan-Andromeda Archaeological Survey, \citealt{McConnachie2009, Martin2016}) are generally well studied.  Typically, these ground-based, resolved stellar studies are limited to the Local Group and work in concert with the extensive number of surveys and studies at larger distances (e.g., M101: \citealt{Karachentsev2019,Muller2017}; Cen A: \citealt{Muller2019,Crnojevic2018}; NGC 1291: \citealt{Byun2020}), in other environments (e.g., NGVS \citealt{Ferrarese2012, Ferrarese2020} and Fornax \citealt{Venhola2018}), and which question the ``stereotypical” nature of the Local Group (e.g., the SAGA Survey: \citealt{Geha2017, Mao2020}; \citealt{Carlsten2020}).

The importance of environment on dwarf galaxy evolution is well known, most famously the presence of increased star formation with distance from a large companion (\citealt{Einasto1974}, also see \citealt{McConnachie2012} and references therein). However, while it is clear that dwarfs around a large host galaxy are affected by their companion, the interplay of these effects remain relatively poorly understood and quantified \citep{Spekkens2014,Geha2012}. The study of dwarfs well separated from large galaxies, where the current impact of the host is likely to be minimal, offer an interesting opportunity. Even here though, we must be aware that some of these dwarfs may be ``backsplash" dwarfs (\citealt{Buck2019} and references therein), which could have had a close pericentric passage with a massive galaxies some point in their history. The isolated dwarfs in the Local Group present a unique opportunity to study the lowest mass galaxies in great detail, largely in the absence of strong environmental influences. 

Of course, isolated Local Group galaxies have been studied for many years. This includes individual dedicated studies for some of the most prominent members e.g., WLM \citep{Wolf1909, Melotte1926, Leaman2012}, IC 1613 \citep{Hodge1991, Pucha2019}, and DDO 210 \citep{McConnachie2006}; as part of catalogues like \cite{Mateo1998a, Karachentsev2013, VanDenBergh1999, McConnachie2012}; as inclusion in HI Surveys like LITTLE THINGS \citep{Hunter2012} or FIGGS2 \citep{Patra2016}; as part of Local Group-wide studies like the star formation history analysis of \cite{Weisz2014}; and as targets to understand their stellar dynamics (e.g., \citealt{Kirby2014, Kacharov2017} among numerous other works). However, homogeneous, systematic studies focused on this population are generally lacking but with exceptions, such as Local Group Galaxies Survey Project (\citealt{Massey2006, Massey2007}). 

Given the limited sample of galaxies available, homogeneous analyses are important in order to minimise systematic errors in what is, fundamentally, a relatively small statistical sample. It is with this intent that we began {\it Solo}, the {\it So}litary {\it Lo}cal Dwarf Galaxy Survey. 

{\it Solo} is a volume limited sample of all known, nearby, and isolated dwarf galaxies, designed to study the secular evolution of dwarfs. A general introduction to the {\it Solo} Survey is given in \cite{Higgs2016}. Briefly, all 44 dwarfs that are known within 3\,Mpc, and which are located more than 300\,kpc from the Milky Way or M31 are included. In this distance range, at least the outer parts of these dwarfs can usually be resolved into stars using ground-based facilities. While the adoption of 300\,kpc as a threshold for isolation is somewhat arbitrary (although it does correspond to the approximate expected halo virial radius for these galaxies; e.g., \citealt{Klypin2002, posti2019}), work by \cite{Geha2012} and \cite{Spekkens2014} suggests that it is around this distance that there is a notable change in dwarf morphology (see also \citealt{Einasto1974}). While there is no doubt that backsplash galaxies will be in the sample (e.g., \citealt{Buck2019} suggest both IC 1613 and And XXVIII are likely backsplash systems), it is clear that most of the {\it Solo} sample of dwarfs have largely evolved in isolation, and are the best nearby systems to use to explore the faint end of the galaxy luminosity function away from the influence of massive galaxies.

In comparison to Local Group Galaxies Survey Project, \cite{Massey2006, Massey2007} focused on detailed analysis of the CMDs using UVBRI photometry of some of the more massive Local Volume dwarfs (M31, M33, NGC 6822, WLM, IC 10, Phoenix, Pegasus, Sextans A, and Sextans B). In contrast, {\it Solo} has a larger field of view by a factor of $\sim 2$ and deeper photometry by $\sim$ 2-3 mags, but in only 2 filters. Our focus is largely on the extended structure of these galaxies rather than their detailed stellar populations, as in \cite{Massey2006, Massey2007}.

The {\it Solo} Survey is designed to probe several questions pertaining to the role of environment in the evolution of the lowest mass galaxies. In particular, the intent is to provide observational data and parameters of a similar quality to those that exist for the Milky Way and M31 satellite populations. This particular contribution is focused towards obtaining updated wide field structural parameters in a similar way to the corresponding M31 and Milky Way data for galaxies in a similar luminosity range, i.e. by analysis of the distribution of the oldest stellar populations, primarily the red giant stars. This is not an insignificant point, given that many of the isolated Local Group galaxies have their light dominated by young stellar populations, which are invariably more centrally concentrated than other stellar populations. \cite{Weisz2014} presents star formation histories showing substantial recent star formation in many of these systems, and studies of radial gradients in the stellar populations of dwarf galaxies invariably show that the younger populations are more concentrated (e.g. \citealt{Harbeck2001,Tolstoy2004, Bernard2008, Mercado2020} among others). Thus, in determining the physical extent of the galaxy, studies that are based solely on integrated light or young stars will generally find a smaller size for these galaxies than studies based on the distribution of older stellar populations (as is nearly always the case for studies of Milky Way and M31 satellites). Minimising these systematic differences between observations of satellites and isolated galaxies is a primary purpose of this current study. 
 
This paper is structured as follows. Section \ref{sec:survey} describes the survey, including the observations, data processing, calibrations, and a brief introduction to the dwarfs analysed.  Section \ref{sec:intlight} analyses the integrated light in the central regions of the target galaxies. Section \ref{sec:basic} describes the analysis of the resolved stellar components, including distance estimates and shape analysis. Section \ref{sec:fit} derives radial profiles from a combined fit of the integrated light and resolved stars, and presents the associated parameters describing the dwarf galaxies. Section \ref{sec:results} compares our results to the literature, and Section~\ref{sec:conclusions} concludes.

\section{Observations and Data Processing}\label{sec:survey}

\subsection{Observations}\label{sec:obs}
The {\it Solo} Survey is a volume limited, wide field imaging survey of all known dwarf galaxies which are within 3\,Mpc and more than 300\,kpc from either M31 or the Milky Way. Galaxies are observed with either CFHT/Megacam in the northern hemisphere or Magellan/Megacam or IMACS in the southern hemisphere. Some targets are observed with multiple instruments for calibration purposes.  The total survey area per galaxy is approximately one square degree, regardless of telescope/instrument (for Magellan, multiple pointings are used to cover this area, whereas only a single pointing is required for CFHT). 
All dwarfs were observed in $g-$ and $i-$ bands. Additionally, almost half (21 out of 44 dwarfs) were also observed in the $u$ band. Some of the $u$ band observations were taken as part of the Canada France Imaging Survey (CFIS; \citealt{Ibata2017}) or are from archival data.

\subsection{The Local Group Subset}

While the full {\it Solo} Survey contains 44 dwarfs, this paper explores a Local Group (LG) subset, consisting of 12 dwarfs within the zero velocity surface of the Local Group ($\approx $1 Mpc) and visible from the northern hemisphere. Table \ref{tab:obsdet} lists these 12 galaxies (hereafter the ``LG subset") and gives details of the observations relevant to this work. 
Other dwarfs within the zero velocity surface but observed for {\it Solo} only from the Southern hemisphere are Cetus, Eri II, Phoenix and Tucana. A full list of the {\it Solo} targets and observational details can be found in \cite{Higgs2016}( hereafter Paper I).

\begin{figure}
	\includegraphics[width=\linewidth]{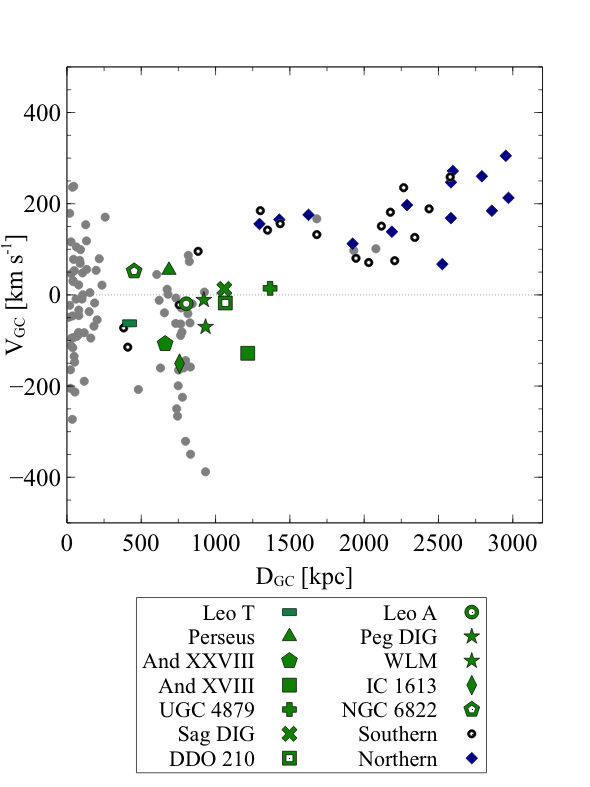}
	\caption{Galactocentric distance (D\textsubscript{GC}) versus radial velocity relative to the Milky Way (V\textsubscript{GC}) of all dwarfs in \protect\cite{McConnachie2012}. {\it Solo} dwarfs with the Local Group subset studied in this paper highlighted in green, with the remainder of CFHT observed $Solo$ dwarfs shown with blue diamonds. Black hollow circles are $Solo$ Local Group dwarfs observed with Magellan and not presented in this paper. Galaxies at large distances in this figure have their velocity dominated by the Hubble flow.  The M31 sub-group is obvious as the cluster of points around $785\pm25$\,kpc \citep{McConnachie2005}.}\label{fig:vhelio}
\end{figure}

Figure \ref{fig:vhelio} shows galactocentric radial velocity against galactocentric distance for dwarfs in and around the Local Group. Our transformation from the heliocentric values listed in \cite{McConnachie2012} assumes ($R_{\odot}$, $V_c$) = (8.122~kpc, 229~km s$^{-1}$) and ($U_{\odot}$, $V_{\odot}$, $W_{\odot}$) = (11.1, 12.24, 7.25)~km s$^{-1}$ (\citealt{Gravity2018}; \citealt{Schonrich2010}). Within this plot, the LG subset can be defined as those whose motion is not affected by the Hubble flow. In contrast, those galaxies further away are clearly affected by the expansion of the Universe. The boundary, the so-called "zero velocity surface", is at approximately 1 - 1.3\,Mpc from the Milky Way (about 1\,Mpc from the center of the Local Group; \citealt{McConnachie2012}). In Figure \ref{fig:vhelio}, the M31 satellite population stands out as the distinct cluster of galaxies around 800\,kpc \citep{McConnachie2005}.

\begin{figure*}
	\includegraphics[width=\linewidth]{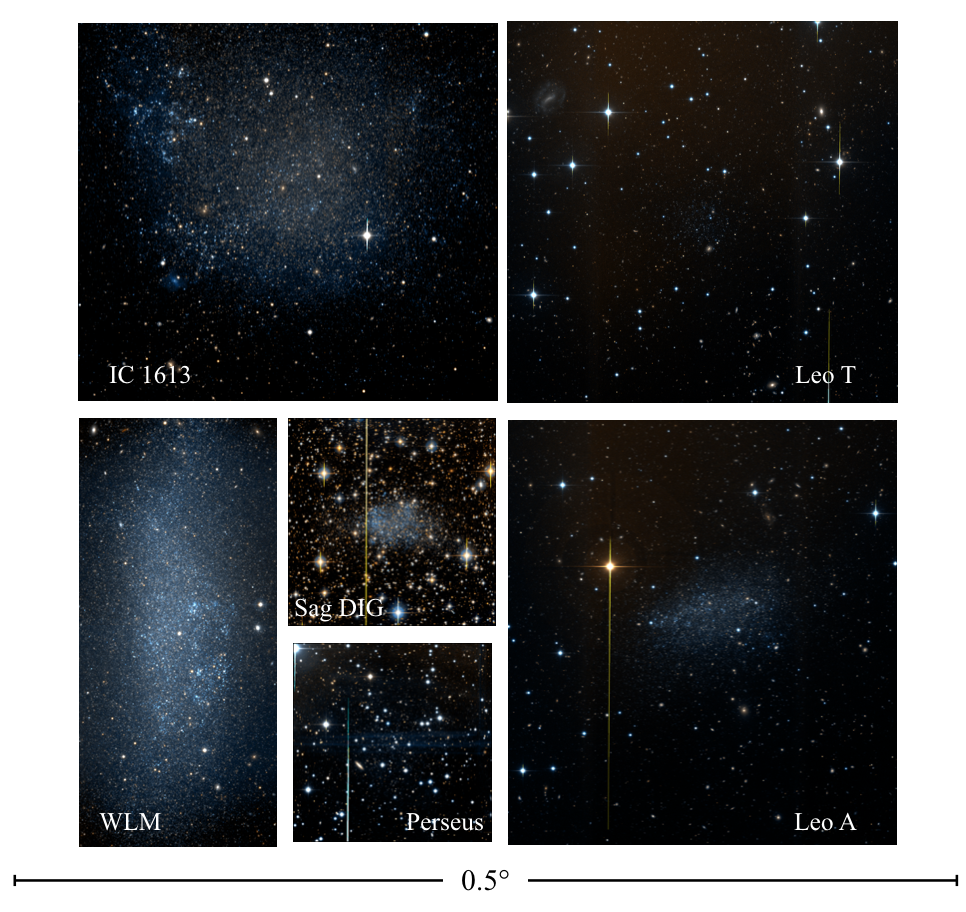}
    \caption{Two - colour images of 6 of the dwarfs in this paper, showing the diversity of  this small sample. The regions shown for each dwarf does not cover the full field of view, but each panel is scaled for the same angular size. The black bar at the bottom of the image is  $0.5^{\circ}$, half of the MegaCam field of view. Each field is oriented with North at the top and East to the left. }
    \label{fig:pretty}
\end{figure*}

 Figure \ref{fig:pretty} shows colour composite images for 6 of the dwarfs in our sample, and qualitatively demonstrates the diversity of the Local Group subset. Striking differences in size, stellar density, number of bright stars, colour, and density of MW foreground can clearly be seen by eye. The dwarfs range from low mass, faint systems like Perseus, to large, obviously star-forming galaxies like WLM. Some dwarfs -- like WLM -- are well known and have been well studied (e.g., \citealt{Leaman2012}, \citealt{Leaman2013} among others),  while others in our sample, like And XXVIII, are less so. 

\begin{table}
	\centering
	\caption{Details on filters used in observations. Those indicated with a $^{\star}$ were taken as part of CFIS.}
	\label{tab:obsdet}
	\begin{tabular}{llccccccc} 
		\hline
        \hline
 		Name &Alt. Name & Filter &Filter ID & Year &  \\
 		\hline
        \hline
         Leo T  &       &  $u$ & u.MP9302 & 2016A$^{\star}$ \\
                          && $g$ &g.MP9402& 2016A &\\
                          && $i$ &i.MP9703& 2016A &\\
         \hline
        Perseus  &  And XXXIII  &  $g$ &g.MP9402& 2016B &\\
         					 && $i$ & i.MP9703& 2016B &\\
        \hline					 
        And XXVIII  &       & $u$ &u.MP9302 & 2017B$^{\star}$\\
                    	     &&  $g$ &g.MP9401& 2012A &\\
                             && $i$ &i.MP9702& 2012A &\\
          \hline
		  And XVIII  &    & $g$ &g.MP9402&  2018B & \\  
           				  && $i$ &i.MP9703&  2018B & \\
          \hline
		  
          UGC 4879 & VV 124 & $g$ &g.MP9402& 2016A &\\
                          &&  $i$ &i.MP9703& 2016A &\\          
          \hline
		  Sag DIG &         & $u$ & u.MP9301 & 2013A & \\
                          &&  $g$ & g.MP9401 & 2012B &\\
                          &&  $i$ & i.MP9702 & 2012B &\\
         \hline
		 DDO 210 & Aquarius & $u$ &u.MP9301& 2013A &\\
                             &&  $g$ &g.MP9401& 2013A \\
                             &&  $i$ &i.MP9702& 2013A \\
        \hline
         Leo A  &         &  $u$ & u.MP9302 & 2015B$^{\star}$ \\
                          &&  $g$ &g.MP9402& 2016A &\\
                          &&  $i$ & i.MP9703& 2016A &\\
         \hline
         Peg DIG & DDO 216   & $u$ &u.MP9301& 2013A &\\ 
                    		 && $g$ &g.MP9401& 2012A &\\
                             && $i$ &i.MP9702& 2012A &\\
         \hline
         WLM    & DDO 221 & $g$ &g.MP9401& 2013B &\\
                          && $i$ &i.MP9703& 2016B &\\
         \hline
         IC 1613 & DDO 8 & $g$ &g.MP9402& 2016B &\\
                          && $i$ &i.MP9703& 2016B &\\
        \hline
		 NGC 6822 & DDO 209 & $u$ &u.MP9301& 2013A \\
                    	  && $g$ &g.MP9401& 2013A \\
                          && $i$ &i.MP9702& 2013A \\
        \hline
        \hline
	\end{tabular}
\end{table}

\subsection{Data processing}

An extensive and detailed description of the data processing is given in Paper I. Preliminary image processing is similar to \cite{Richardson2011}. The preprocessing (de-biasing, flat-fielding and fringe corrections) is completed using CFHT's Elixir system prior to transferring the data to the Cambridge Astronomical Survey Unit. Here, the data is reduced using the adapted pipeline originally developed for the Wide Field Camera observations from the Isaac Newton Telescope (see \citealt{Irwin1985, Irwin1997}, \citealt{Irwin2001} and \citealt{Irwin2004} for details).

Catalogues of stellar sources, matched in each band ($g$, $i$ with some additional $u$ band) are generated for each dwarf, although in what follows, we focus exclusively on $g$ and $i$ observations. Before stacking all exposures, catalogues are generated for each science image, which determines the data quality and refines the astrometric calibration. For details, see Paper I and references therein. Each science exposure is assessed based on the average seeing, ellipticity of point sources, sky level and sky noise. The images are stacked using the World Coordinate System (WCS), and the source catalogues generated for each exposure to align the images with a sub-pixel precision. Overlapping background regions are used to account for sky variations between exposures in the stack. Images in the final stack were weighted in accordance to the average seeing. Cosmic rays were removed. Source catalogues are generated from these full stacks and the WCS astrometry is re-derived. The sources are classified using their morphology as stellar or non-stellar and a final catalogue including both bands is produced. If two sources are identified in the different bands and lie within 1" of each other, they are considered the same object, and the position for the higher signal to noise object is retained. Any object which is detected in only one band is retained as the catalogues are merged. For some dwarfs, seeing for one band (generally $i-$) was significantly better than the other band. In this situation, sources detected in the better band were solely used to generate the final catalogue using forced photometry for the lower quality band. 

\subsubsection{Photometric Corrections and Calibrations}

The CFHT filter system changed during the period of time in which the observations were taken. (see Table \ref{tab:obsdet} for details). As such, we transform all the filters to the current CFHT MegaCam filter system. The relevant filters in this current system are identified as $u$ (MP9302), $g$ (MP9402) and $i$ (MP9703). The previous CFHT filter system, $uS$ (MP9301), $gS$ (MP9401) and $iS$ (MP9702), was in use until 2015. The older system are converted to the new filters using Equations \ref{convert1} and \ref{convert2} (S. Prunet, private communication): 

\begin{eqnarray}
g - gS= 0.022 + 0.078 (gS - iS) \label{convert1} \\
i - iS = 0.001 - 0.0015 (gS -iS) \label{convert2}
\end{eqnarray}

\noindent The magnitudes are presented in their natural instrumental (AB) system. The $g-$ and $i-$ magnitudes of each source is corrected for Galactic dust using values interpolated from the \cite{Schlegel1998} dust extinction maps, assuming that $A_g = 3.793 \times E(B-V)$ and $A_i=2.086 \times E(B - V)$.
No corrections are applied for extinction internal to the dwarfs. 

Only stellar sources are used for the subsequent analysis, using the morphological classification procedures described in \cite{Irwin1985,Irwin1997}. Sources within 2 $\sigma$ of the stellar locus in both bands are considered stellar. From the source catalogues, the positions of the sources are de-projected, converting right ascension (R.A.)  and declination (Dec.) to standard plane coordinates ($\xi,\eta$). The standard plane is centered on the dwarf's coordinates as listed in \cite{McConnachie2012}. 

\begin{figure}
	\includegraphics[width=\linewidth]{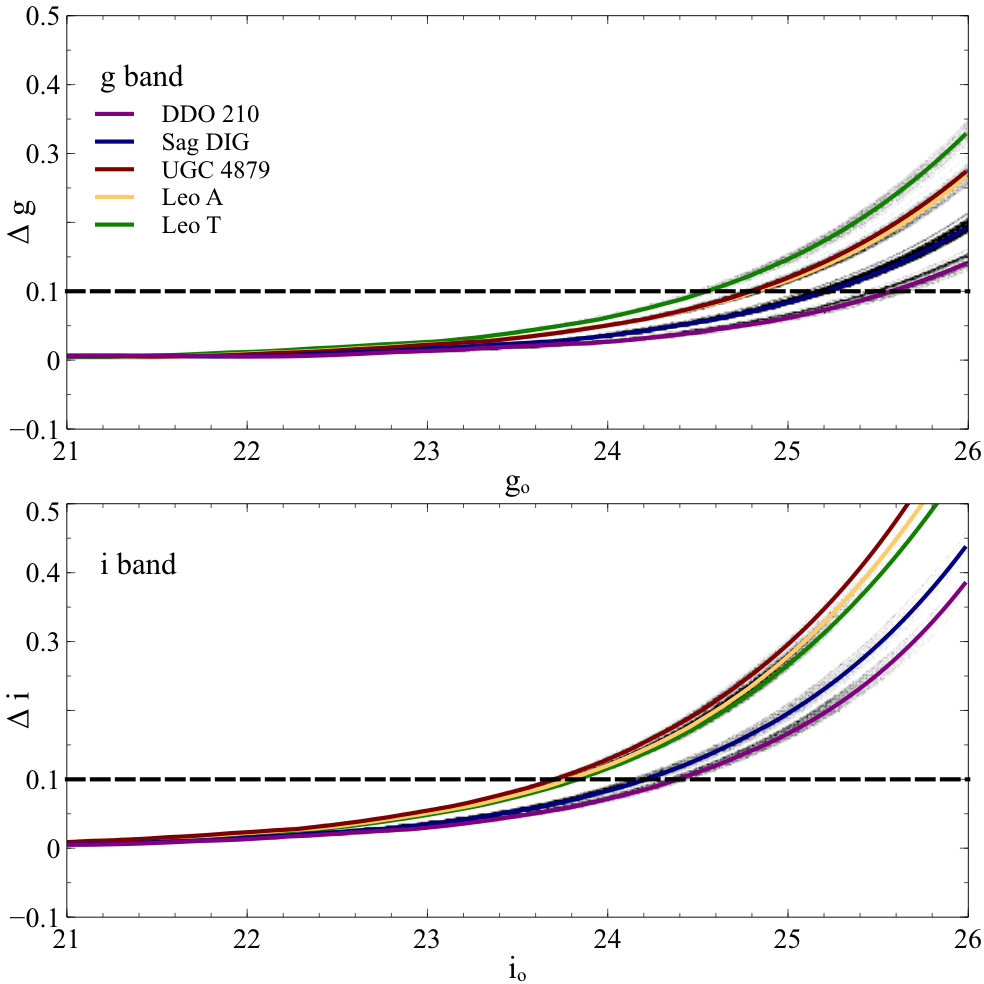}
    \caption{ Representative photometric errors as a function of $g$- ({\it upper panel}) and $i$-band ({\it lower panel}) for DDO 210, Leo T, Leo A, Sag DIG and UGC 4879. The grey shaded background shows a histogram of all points.}
    \label{fig:photerrs}
\end{figure}

Figure \ref{fig:photerrs} shows photometric errors as a function of magnitude in $g$ and $i$ for a sample of dwarfs. These uncertainties are derived from the square root of the variance of the weighted flux measurements. They are generally less than 0.1 magnitudes at $(g \sim 24.5, i\sim 24.0)$, meaning red giant branch (RGB) stars in the dwarfs are easily detectable in all cases.

WLM was observed in non-photometric conditions for the $i-$ band. As such, a correction was required and was derived from  Isaac Newton Telescope (INT) Wide Field Camera observations of WLM, DDO 210 and Peg DIG (published in \citealt{McConnachie2005}). RGB stars in common between the two datasets for DDO 210 and Peg DIG were used to define the appropriate transformation between the two filter systems for photometric conditions. Applied to the INT data for WLM, this transformations then allowed us to determine the appropriate correction to convert the non-photometric MegaCam data to the correct photometric system.

 \subsection{Summary of target galaxies}
 
Here, we briefly summarise some of the notable properties of each of our target dwarfs, and highlight the pertinent literature as it relates to their structural properties. This background is not intended to be an exhaustive summary of all previous work, but rather a useful and easy reference for the interested reader. 

\subsubsection{Andromeda XVIII}

 And XVIII is in the vicinity of M31, and was discovered by \cite{McConnachie2008} as part of precursor observations for the Pan-Andromeda Archaeological Survey (PAndAS; \citealt{McConnachie2009}). Unfortunately, in these discovery observations, the majority of the galaxy fell in one of the large chip gaps on the detector, obscuring the central region. The dwarf's parameters were estimated under the assumption of circular symmetry to address the obscured portion. \cite{Martin2016} analysed the dwarf as part of a wider study of M31's dwarfs, however they also used the same data. They estimated its structural parameters, and included a more robust surface brightness estimate, however its ellipticity was poorly constrained. Here, new observations, placing the dwarf well away from chip gaps, have been obtained which allow for a complete picture of its global structure.

\cite{Tollerud2012} obtained kinematics of And XVIII stars as part of the SPLASH survey. They concluded that it is a kinematically cold, dark matter dominated dwarf. Given its distance from M31 (of order 600\,kpc distant), it is unlikely to have interacted with this massive galaxy, but \cite{Tollerud2012} noted that there is a remarkable agreement between their systematic velocities. 

\subsubsection{Andromeda XXVIII}
 \cite{Slater2011} determined the structural parameters for And XXVIII from SDSS observations using the maximum likelihood methods described in (\citealt{Martin2008};  our new observations are approximately 3 magnitudes deeper than the SDSS observations). \cite{Slater2015} followed up with Gemini/GMOS and Keck/DEIMOS observations, finding no signs of recent star formation but the presence of a red clump and other features suggested an extended period of star formation. \cite{Buck2019} suggest that this dwarf has a highly likelihood (91\%) of having interacted with the MW (i.e. it could be a backsplash galaxy). This estimate is based on the comparison of observed galaxy properties (distance, radial velocity and velocity dispersion) with simulated satellite and field dwarfs in the NIHAO simulations (\citealt{Wang2015}). 

\subsubsection{The Pegasus Dwarf Irregular Galaxy}
 Peg DIG (also known as DDO 216 and UGC 12613) is a transition-type dwarf that is host to a large, central globular cluster \citep{Cole2017,Leaman2020}, making it among the lowest mass dwarfs in the Local Group with a globular cluster (in addition to Eri II; see \citealt{Zoutendijk2020} and references therein). A notable population of RR Lyrae stars in the cluster have been used to derive a distance of  $24.77\pm 0.08$ \citep{Cole2017}. 

\cite{McConnachie2007} found an asymmetry between the gas and stellar components indicative of ram pressure stripping removing the gas from the system. However, more recent HI observations by \cite{Kniazev2009} contradict this conclusion, finding minimal evidence of interactions. These authors concluded that this dwarf has only recently joined the Local Group.  The structural parameters derived by \cite{Kniazev2009} provided a good comparison to this current work, however the SDSS observations used are significantly (> 3 magnitudes) shallower than the {\it Solo} data.  They found the dwarf is more extended than previously thought and noted possible stellar extensions on the north-west end. In their HI observations, \cite{Kniazev2009} observed solid body like rotation. \cite{Kirby2014} found rotation in the red giant branch stars as well. 

\subsubsection{Leo T}

Leo T is one of lowest mass and closest dwarfs in {\it Solo}. It is notable for being one of the few very faint dwarfs with a young stellar population, and like Peg DIG, it is often described as a transition-type dwarf, given that its other characteristics are generally typical of dwarf spheroidal galaxies.  \cite{DeJong2008} determined its structural parameters based on deep, wide-field Large Binocular Telescope data, and it is also included in the MegaCam Survey of Outer Halo Satellites \citep{Munoz2018b,Munoz2018a}, using CFHT $gr$ photometry. Its HI structure was most recently studied by \cite{Adams2018}. \cite{Teyssier2012} argued that both Leo T and NGC 6822 have likely had a close pericentric passage with the Milky Way, whereas \cite{Buck2019} suggested that there is only an approximately 50\% chance of Leo T being a backsplash galaxy. 

\subsubsection{Leo A}
Leo A (DDO 69) shows delayed star formation, like DDO 210, and only minimal star formation happened before 10 Gyrs ago. As such, it is sometimes argued as being a possible survivor of reionization \citep{Cole2007,Cole2014}. \cite{Stonkute2014,Stonkute2018} presented detailed studies of Leo A based on Subaru/SuprimeCam data, in addition to using HST ACS fields at large radius. \cite{Stonkute2015} reported the discovery of a young star cluster in Leo A. \cite{vanzee2006} studied the HII regions in Leo A, confirming that its interstellar medium has one of the lowest metallicities of nearby galaxies (\citealt{vanzee2006} find 12 + log(O/H) = $7.30\pm0.05$). This low metallicity makes it a useful system for calibration (e.g. for Cepheid observations \citealt{Bernard2013}). Leo A was also used by \cite{ruizlara2018} to test whether there are differences between star formation histories derived from colour magnitude diagrams compared to integrated light studies. 

\subsubsection{Perseus}
Perseus I, also known as And XXXIII, was discovered by \cite{Martin2013b} in the Pan-STARRS1 3$\pi$ survey and followed up \cite{Martin2014} with spectroscopy. The structural parameters derived are based on the maximum likelihood method \citep{Martin2008}. The {\it Solo} observations are about 2 magnitudes deeper than the PanSTARRS data. The location of this dwarf makes it an interesting object with respect to the ``planes of satellites" around the Local Group (see \citealt{Pawlowski2014} for a discussion of Perseus in this context).

\subsubsection{UGC 4879}

UGC 4879 (also known as VV 124) was first studied in detail by \cite{Kopylov2008}. \cite{Bellazzini2011a,Bellazzini2011b} performed a comprehensive analysis using HST ACS, the Large Binocular Telescope, and HI observations, and identify 2 young star clusters. \cite{Kopylov2008} and \cite{Jacobs2011} noted that it is a very isolated dwarf, and so likely not a backsplash system. 
Its structure is somewhat unusual, insofar as there are two ``wings" on either side of an otherwise  elliptical system, perhaps suggestive of a stellar disk. \cite{Kirby2012} placed an upper limit on the possible stellar disk rotation of $8.6$\,km s$^{-1}$.

\subsubsection{DDO 210}

DDO 210 (also known as Aquarius) is classified as a transition type dwarf in that there is no current star formation observed but, prior to the discovery of Leo T, it was the faintest known dwarf with a significant gaseous component. Along with Leo A, DDO 210 has a delayed star formation history \citep{Cole2014}, meaning that most of the stars are of intermediate age. \cite{Cole2014} noted a strong blue horizontal branch population which is unusual relative to similar dwarfs (like Leo A), as well as the presence of a red clump and a red bump. Its free fall time is longer than a Hubble time, hence it is likely on its first in-fall into the Local Group  (\citealt{McConnachie2012}). It is well separated from neighbouring galaxies. 

The HI and stellar components show some offset and differences in spatial structure \cite{McConnachie2006}. \cite{HermosaMunoz2020} looked at the metallicity and kinematics of stars, finding significant misalignment of the HI and stellar components. They suggested that this may be due to recent HI accretion or re-accretion of gas after feedback from star formation. 

\subsubsection{The Sagittarius Dwarf Irregular Galaxy}

Sag DIG was studied as the first {\it Solo} dwarf in Paper I. We previously noted differences between the HI structure and its stellar structure, similar to DDO 210 (Aquarius).  There are faint extended RGB stars regions along the semi major axis of this dwarf. Qualitatively, these extensions are not found to have a HI counterpart in LITTLE THINGS observations \citep{Hunter2012}. Sag DIG lies at a low galactic latitude so the foreground contamination from the MW in the field is significant.


\subsubsection{Wolf-Lundmark-Melotte}

WLM, also known as DDO 221, is one of the more massive dwarfs in the {\it Solo} Survey. It has been extensively studied, from its discovery over 100 years ago \citep{Wolf1909} to recent work presented in \cite{Albers2019}. Several spectroscopic studies have been conducted. For example, RGB stars have been targeted by \cite{Leaman2009, Leaman2012} and blue supergiants have been targeted by \cite{Venn2003, Bresolin2006, Urbaneja2008}.
\cite{Leaman2012} concluded that this dwarf is a highly oblate spheroid. They found good agreement, kinematically and photometrically, between the position angle and ellipticity of the RGB stars, young carbon stars and HI gas. They concluded that external environmental effects have not had a dominant role in shaping the structure of WLM. 
Using HST observations, \cite{Albers2019} derived a SFH history and found relatively continuous star formation, with no dominant early episode (\citealt{Gallart2015} classified the SFH of WLM as "slow"). 
This dwarf is one of the few with a known globular cluster, in this case a single large globular cluster located just outside the central region.

\subsubsection{ NGC 6822 }
NGC 6822 is one of the more massive dwarfs in {\it Solo} and has been extensively studied. \cite{Demers2006} identified this dwarf as a polar ring galaxy, based on kinematics of HI compared to asymptotic giant branch stars. This claim is disputed by \cite{Thompson2016} who found that the axis of rotation for the stars and HI are largely aligned. \cite{Battinelli2006} derived the stellar structure from wide-field CFHT observations, and noted that it has a bright central bar. NGC 6822 hosts a population of globular clusters \citep{Huxor2013, Veljanoski2015}. \cite{Hwang2014} used extended star clusters as kinematic tracers and derived a mass to light ratio of $ 75^{+45}_{-1}(M/L)_{\odot}$, making NGC 6822 dark matter dominated. 
\cite{Teyssier2012} suggested NGC 6822 is possibly a backsplash galaxy, based on its current observed position and radial velocity. In contrast, McConnachie et al. ({\it submitted}) find NGC 6822 is highly likely to be isolated and on its first infall, based on orbital properties derived using Gaia Data Release 2 proper motions of the brightest stars (\citealt{Gaia2018}). 

\subsubsection{ IC 1613 }
IC 1613 is another massive dwarf in {\it Solo} and has been well studied, including  by \cite{Bernard2007} and recently by \cite{Pucha2019}. \cite{Pucha2019} found strong age gradients, and the young stellar population has a considerably smaller scale radius than the older stars. The SFH they determined is consistent with this picture, with a decreasing star formation rate with increasing radius. They found that the young population is off center, and the star forming regions are distributed in a ring, coincident with areas of high HI column densities. \cite{Buck2019} suggests that this dwarf, along with And XXVIII, has a highly likelihood of having previously interacted with the Milky Way.

\section{Characterizing {\it Solo} Dwarfs using Integrated Light}\label{sec:intlight}

The {\it Solo} Survey selects dwarfs such that at least their outer regions can be resolved into stars from the ground in reasonable seeing. In the central parts of most of these dwarfs, crowding becomes a significant issue and the stellar catalogues become increasingly incomplete.  As a result, these central regions are better analysed using the integrated light rather than resolved stars, particularly with respect to their radial profiles in both bands (and which allows identification of any colour trends present).  

\begin{figure}
\begin{center}

	\includegraphics[width=.9\linewidth]{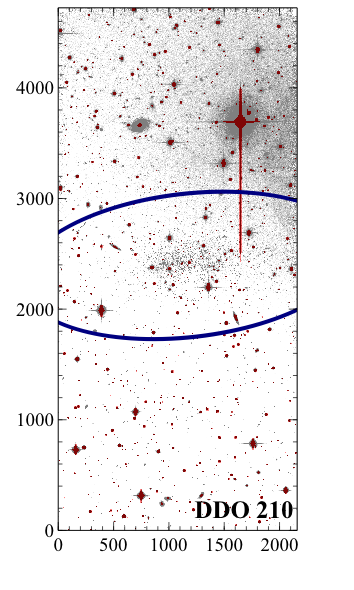}
    \caption{The central chip in the $i-$band for DDO 210. The grey scale shows the integrated flux, scaled to highlight the brighter stars in the dwarf. The red regions are masked out when determining the integrated light profile. The blue ellipse shows the inner edge of the region used to determine the background in the integrated light. The relative inhomogeneity  of the background as well as the presence of very bright stars is obvious and highlights the necessity of carefully selecting the background region.}
    \label{fig:intex}
   \end{center}
\end{figure}

\begin{figure}
	\includegraphics[width=\linewidth]{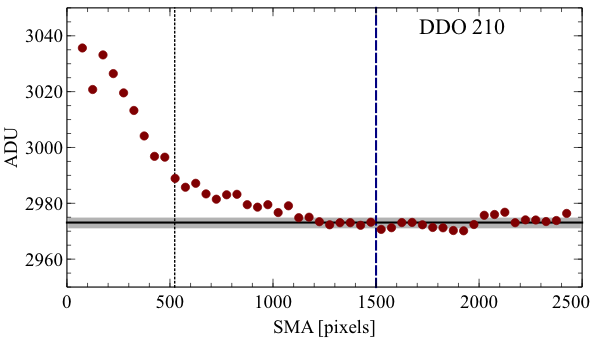}
    \caption{The radial (SMA - semi major axis) profile in ADUs for DDO 210 in the $i-$band. The black line indicates the mean background and the standard deviation  is shown as the grey shaded region. This background is determined using the points outside the blue dashed line (corresponds to the blue ellipse in \ref{fig:intex}). The black dotted line is $R_s$, as defined in Section \ref{sec:fit}.}
    \label{fig:aduex}
\end{figure}

The integrated light profile is determined in both the $g-$ and $i-$ bands. Here, elliptical annuli with the median position angle and ellipticity as derived from the resolved RGB stars (see next section) is used.  All bright sources and their surrounding pixels are masked to reduce contamination from other sources, including saturated objects and diffraction spikes. Figure \ref{fig:intex} shows an example of an $i-$ band image and the applied mask for DDO 210. Note, the central regions of each galaxy fit within a single CCD, and so therefore we only do the integrated light analysis for a single CCD per galaxy. This helps to minimise systematic uncertainties due to detector responses and other instrumental effects.
 
To estimate the background, the median luminosity is found within annuli that are well separated from the galaxy i.e., that are outside of a large ellipse centered on the galaxy. The size of this ellipse varies from galaxy to galaxy from 2 to 4~$R_s$.  The inner bound of this background region is shown in Figure \ref{fig:intex} for DDO 210 as the blue ellipse. To determine the uncertainty on the background estimation, we look at the variation between annuli. This technique accounts for the non-uniformity of the background and acknowledges that systematic errors are dominant compared to Poisson uncertainties. These systematic errors have spatial scales associated with them, depending on the source, and so it is important that the annuli we use to measure these variations are the same width as are used for the main profile. For example, some galaxies like Leo A have a bright star in the image. Consequently, internal reflections within the MegaCam instrument can create large halos, hence the background across the image is not expected to be uniform.  The radial profile of DDO 210 is shown in ADUs in Figure \ref{fig:aduex}, along with the background estimate and the adopted uncertainty. 
 
\begin{figure*}
	\includegraphics[width=.9\linewidth]{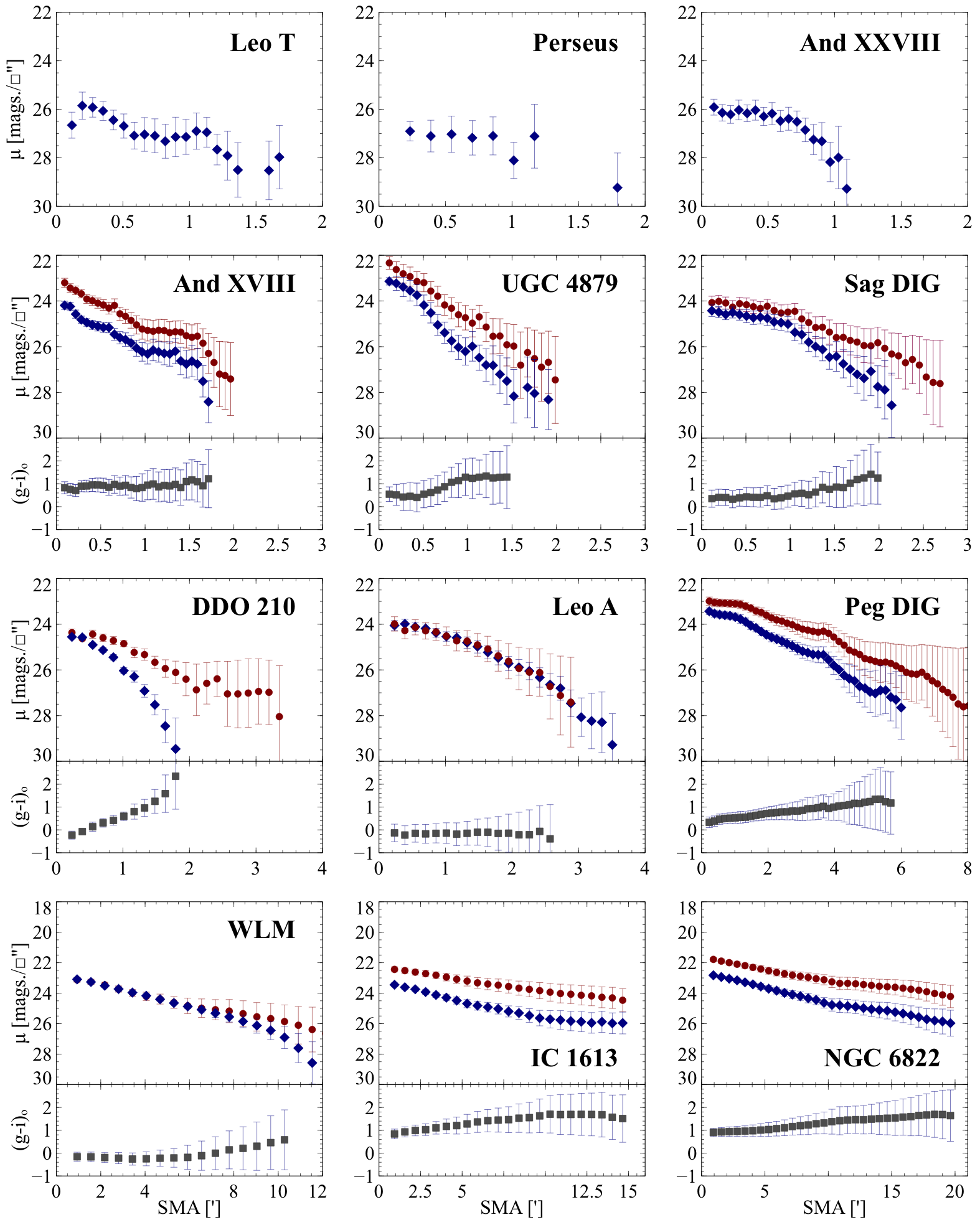}
    \caption{Each panel shows the surface brightness profile (shown per square arc minute - $\square ''$)  from the integrated light in the $g-$ (blue diamonds) and $i-$ (red circles) bands as a function of radius (SMA - semi major axis).  In 3 cases there is no discernible $i-$band profile (top row). The sub-panels below each main panel shows the colour profile for all cases where both $g-$ and $i-$band profiles exist.}
    \label{fig:intprofs}
\end{figure*}

The background subtracted profiles for all {\it Solo} dwarfs are shown in Figure \ref{fig:intprofs}. For some of the faintest systems, there is no discernible $i-$band integrated light component above the background. We show the colour profiles as sub-panels to the main panels for those dwarfs where both $g-$ and $i-$band profiles were determined.

For all nine galaxies with observable colour profiles, the colour profiles are largely flat or slightly rising, indicating blue centers and increasingly red toward larger radii. This structure is consistent with the fact that the {\it Solo} dwarfs are dwarf irregulars, with recent or ongoing star formation. 

\section{Characterizing {\it Solo} Dwarfs using Resolved Stars}\label{sec:basic}

 \subsection{Colour magnitude diagrams and distance estimates}\label{semd:cmds}

\begin{figure}
	\includegraphics[width=\linewidth]{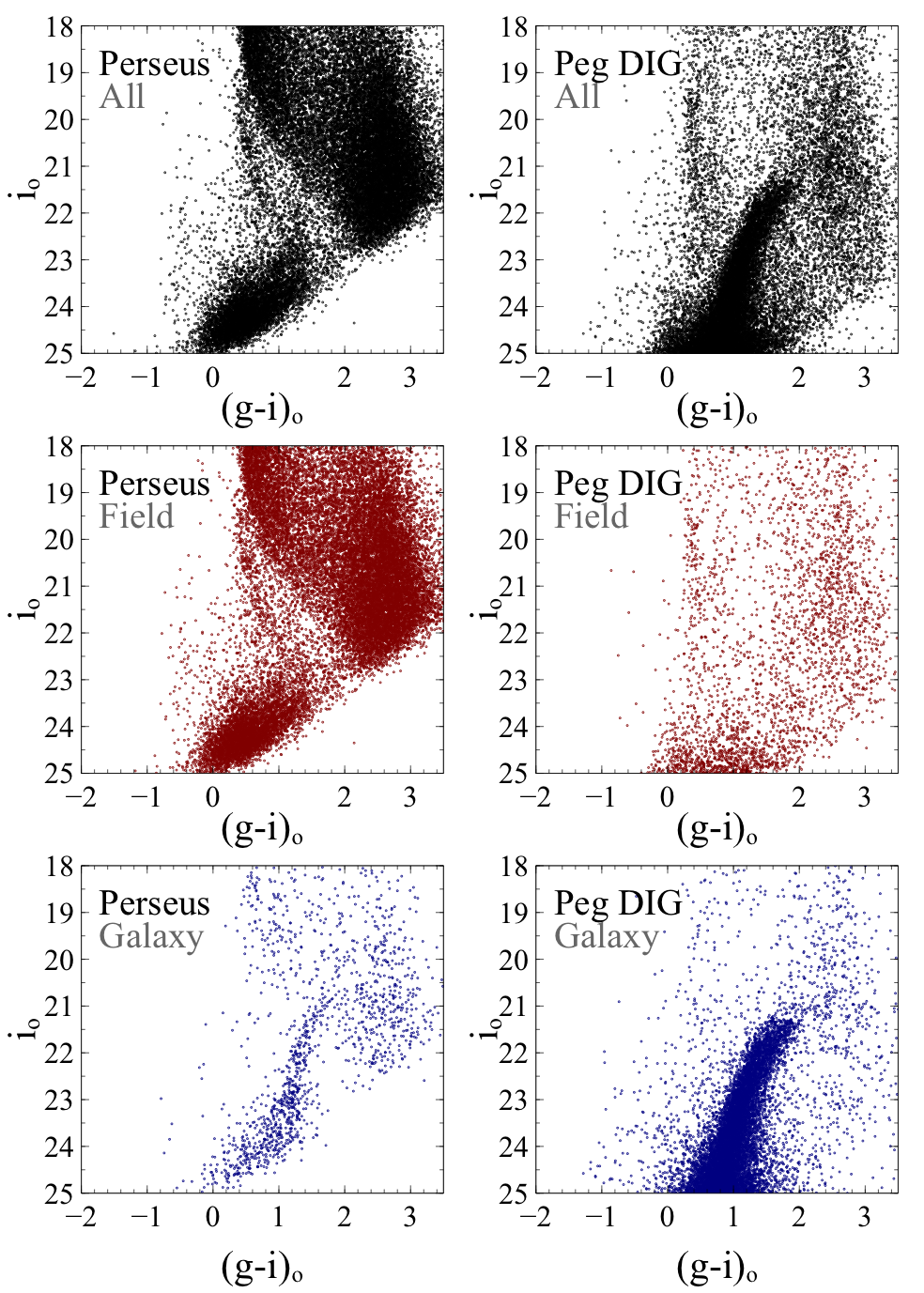}
    \caption{ Examples of contamination in the CMDs of {\it Solo} dwarfs. \textit{Top panels:} All stars in the full 1 sq. degree field of view for Perseus ({\it left}) and Peg DIG ({\it right}). \textit{Middle panels:} All stars in the field far ($> 5 x R_s$) from the dwarf galaxy, showing the different levels of foreground contamination from MW stars, as well as the different structure of the foreground populations. Bright stars around $(g-i)_o=0.5$ are Milky Way main sequence turn-off stars. Stars in the large cloud of points around $(g-i)_o \sim 2.5$ are generally low mass Milky Way main sequence stars. For Perseus, the sequence of stars extending from the main sequence turn-off region to $(0.5, 23)$ is a Milky Way halo overdensity. At faint magnitudes, increasing numbers of background galaxies are misidentified as stars. \textit{Lower panels:} All stars in the field close to the centers of the galaxies, clearly showing the RGB of the two galaxies.}
    \label{fig:bgcmdex}
\end{figure}

Figure \ref{fig:bgcmdex} shows the colour magnitude diagrams (CMDs) of two {\it Solo} galaxies to illustrate the different degrees of contamination from the Milky Way foreground. Perseus is relatively low apparent luminosity ($m_V = 14.1 \pm 0.7$) at a relatively low Galactic latitude ($b = 147.8^{\circ}$), whereas Peg DIG is brighter ($m_V = 12.6 \pm 0.2$) at a higher latitude ($b = 94.8^{\circ}$). Note, in what follows, the term "background" is used to describe foreground contamination arising from the Milky Way as well as from true background sources (generally distant, compact galaxies misidentified as stars at faint magnitudes). In both cases in Figure \ref{fig:bgcmdex}, the stellar populations of the dwarf are most easily discerned by selecting only stars near the center of the galaxy. 

\begin{figure*}
	\includegraphics[width=\linewidth]{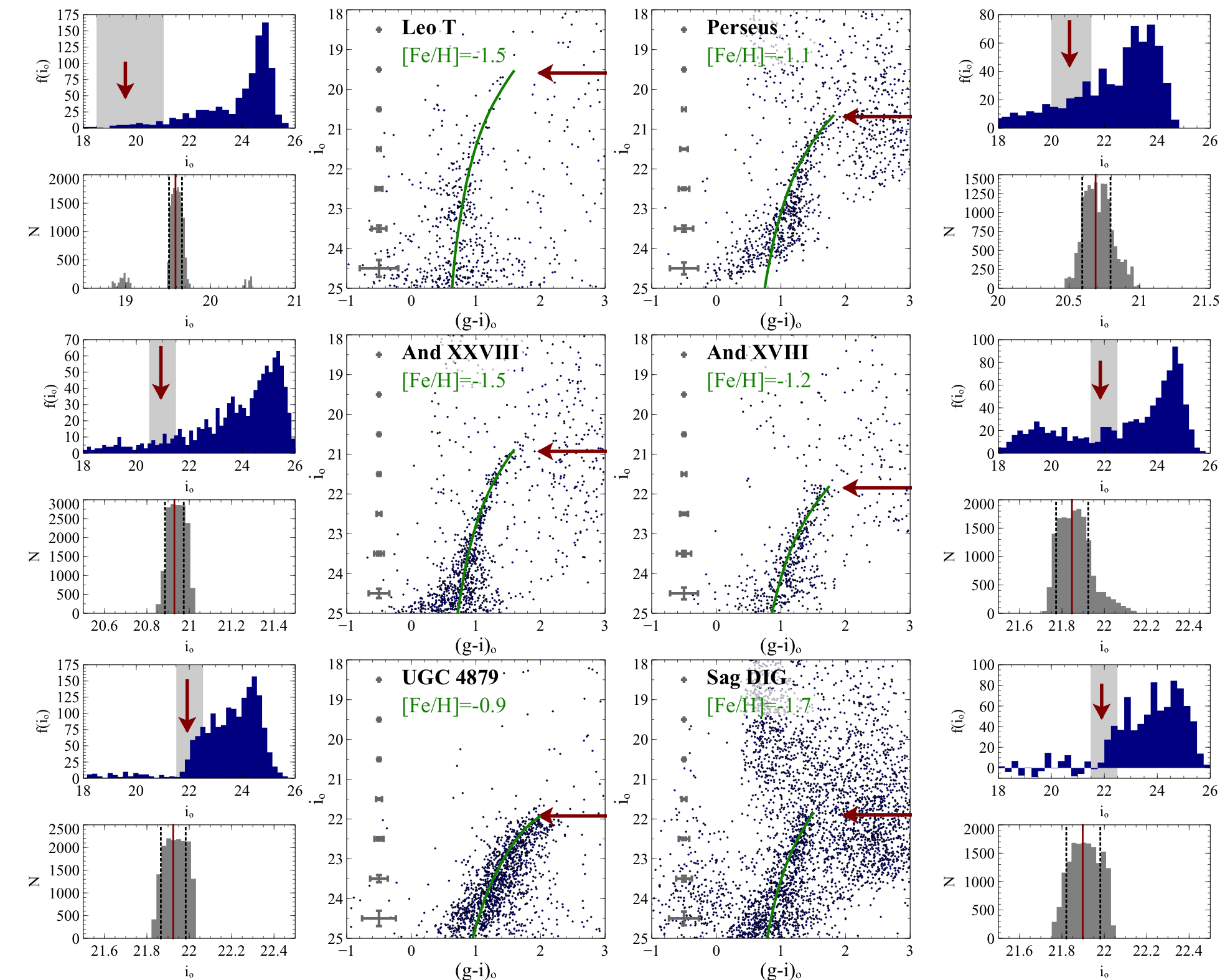}
    \caption{ CMDs and luminosity functions for each of the dwarfs. The large panel for each galaxy is the CMD for stars selected  to be within 2 $R_s$ of the galaxy center. Mean errors as a function of magnitude are shown, and the TRGB is indicated with the red arrow.  The upper small panel is the luminosity function for the RGB, and the lower small panel is the resulting distribution of the measured position of the TRGB (see text for details). The grey shaded region in the upper panel indicates the range shown the lower panel.  Here, the red solid line indicates the median value and the dashed lines show a 16\% and 84\% quartile range (approximating $1 \sigma$).}
    \label{fig:cmds_1}
\end{figure*}

\begin{figure*}
	\includegraphics[width=\linewidth]{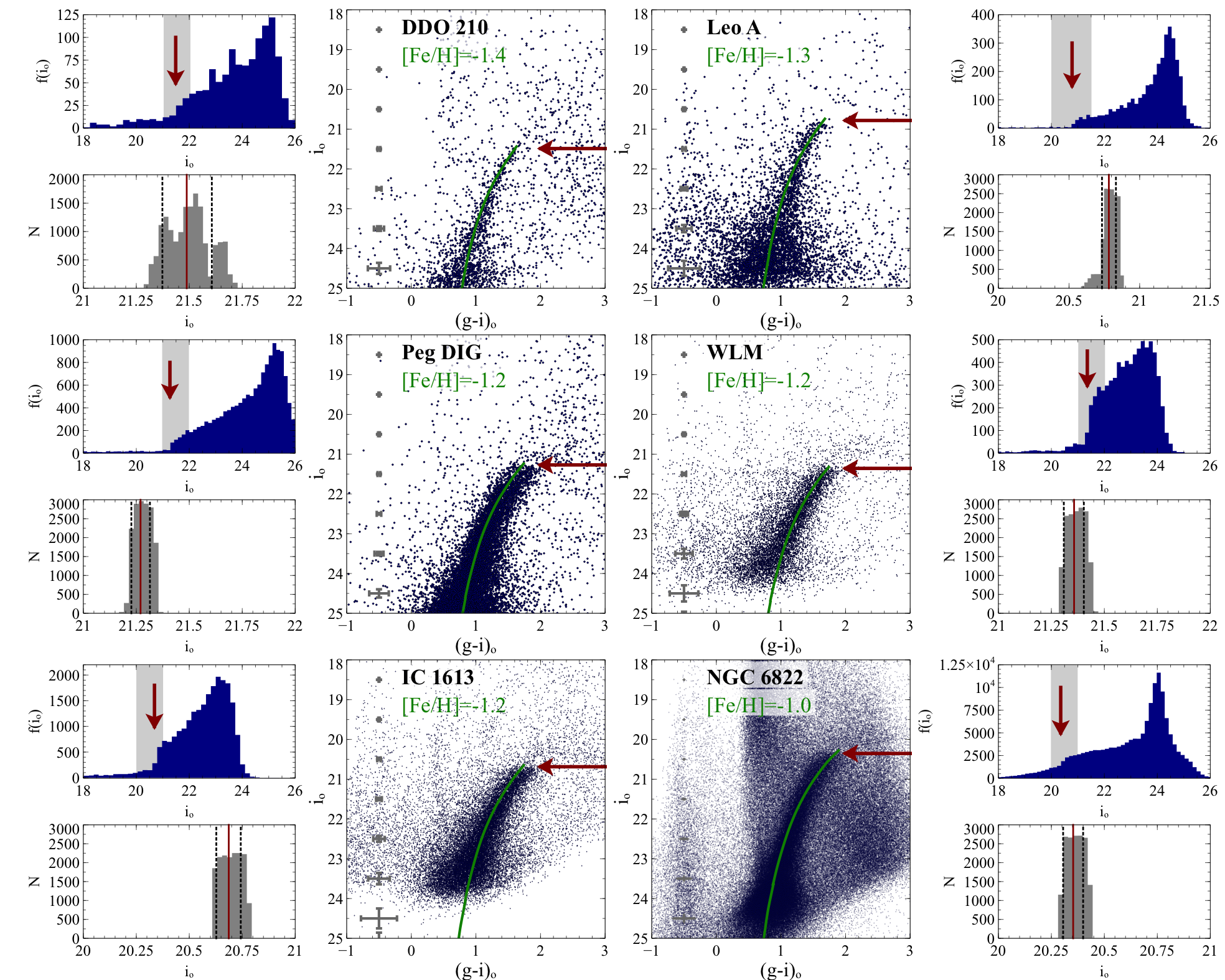}
    \caption{The same as Figure \ref{fig:cmds_1}, CMDs and luminosity functions for each of the dwarfs. The large panel for each galaxy is the CMD for stars selected  to be within 2 $R_s$ of the galaxy center. Mean errors as a function of magnitude are shown, and the TRGB is indicated with the red arrow.  The upper small panel is the luminosity function for the RGB, and the lower small panel is the resulting distribution of the measured position of the TRGB (see text for details). The grey shaded region in the upper panel indicates the range shown the lower panel.  Here, the red solid line indicates the median value and the dashed lines show a 16\% and 84\% quartile range (approximating $1 \sigma$).} 
    \label{fig:cmds_2}
\end{figure*}

The CMDs for all of our sample are shown in Figures \ref{fig:cmds_1} and \ref{fig:cmds_2}. The main feature visible in all CMDs are RGB stars. Young, blue stars are also present (for those with recent star formation), and can be seen as vertical sequences around $(g - i)_o \simeq -0.5$. Sag DIG, Leo A, WLM, IC 1613 and NGC 6822 show a clear blue sequence, where as DDO 210, Peg DIG (and tentatively Leo T and Perseus) show a few stars which may be young, blue stars. Foreground contamination makes distinguishing individual stars associated with the dwarfs not possible in this region without additional observations. Asymptotic giant branch (AGB) stars may also be present for some of the galaxies, although these are hard to distinguish from RGB stars, except at brighter magnitudes, where they are brighter than the tip of the RGB (e.g., WLM). On each panel, a Dartmount isochrone \citep{Dotter2008} is shown, adjusted to the correct distance using our tip of the RGB measurements (see below) and with a metallicity selected to match the average colour of the RGB. The metallicities are listed in the upper right of each plot, ranging from [Fe/H] = -0.9 (UGC 4879) to -1.7 (Sag DIG). This range is broadly similar to the satellites of the Milky Way and M31 with the satellites tending to have additional lower metallicity dwarfs (from values listed in the updated tables of \citealt{McConnachie2012} and respective references). The isochrones plotted are selected to be old ($\sim 12$ Gyrs) and intended to help visualize the RGB. As these dwarfs are irregulars, they are not necessarily simple, old stellar populations. For instance, the Perseus is not ideally matched, however this does not impact the distance estimate.

The distance to each dwarf is found using the tip of the RGB (TRGB) as a standard candle. The TRGB is ideal as its luminosity is insensitive to small variations in age or metallicity given an older, metal poor population \citep{Lee1993}. Indeed, using the Darthmouth isochrones \citep{Dotter2008}, the difference between a 10 Gyrs, 11 Gyrs or 12 Gyrs stellar population is on the order of $\Delta i = 0.002$. This uncertainty is not significant given that the uncertainties in identifying the TRGB from the CMD are on the order of $\Delta i = 0.1$. 

First, the RGB stars are selected from a CMD using parallel ``tram-lines". These tram-lines are visually selected and generously encompass the RGB branch.
If the RGB branch is well populated, the TRGB is apparent by a significant ``break" in the RGB luminosity function. The location of this break is identified using a 5 point Sobel filter, similar to \cite{Sakai1996}. As the luminosity function is binned, we remove the dependence of the choice of bin locations and widths by repeating the analysis with different binning choices, similar to the method used by \cite{Battaglia2012}.  Note that we adopt upper and lower bounds on the region in which the TRGB may be found, to avoid spurious signals due to structures like the red clump or random stars clearly unassociated with the TRGB. These limits are chosen visually and generous encompass the whole region in which the TRGB is located. For low mass dwarfs with poorly populated RGBs, the limits are selected cautiously as to avoid biasing results.  
The distribution of resulting TRGB values (10\,000 combinations of bin width and location in total) is shown with the median values and $1 \sigma$ interquartile range highlighted.

In the above analysis, we do not find it necessary to subtract a reference luminosity function (due to the Milky Way contamination), except in the case of Sag DIG. Typically, the contamination is fairly uniform near the TRGB and due to the low number of stars in some galaxies, subtracting a background did not help in identifying the TRGB, but rather just added more noise. However, for Sag DIG, which has a reasonably well populated RGB branch and substantial contamination we found it was necessary to statistically remove this background. As detailed in Paper I, a portion of the foreground structure in this direction is likely due to the Sagittarius stream. 

The distance modulus, $(m-M)_o$, is computed from the TRGB using an absolute magnitude of $i_{o,MP9703} = -3.49$. This value is derived from the Darthmouth isochrones \citep{Dotter2008} using the median value for old (12 Gyrs) stellar populations with [Fe/H] \textless 0.0 (consistent with the isochrones matched previously). The TRGB value is converted to the new CFHT filters using the previous equations (Eqns. \ref{convert1} \& \ref{convert2}).

\subsection{Structural analysis}\label{sec:stucture}

\begin{figure*}
	\includegraphics[width=.75\linewidth]{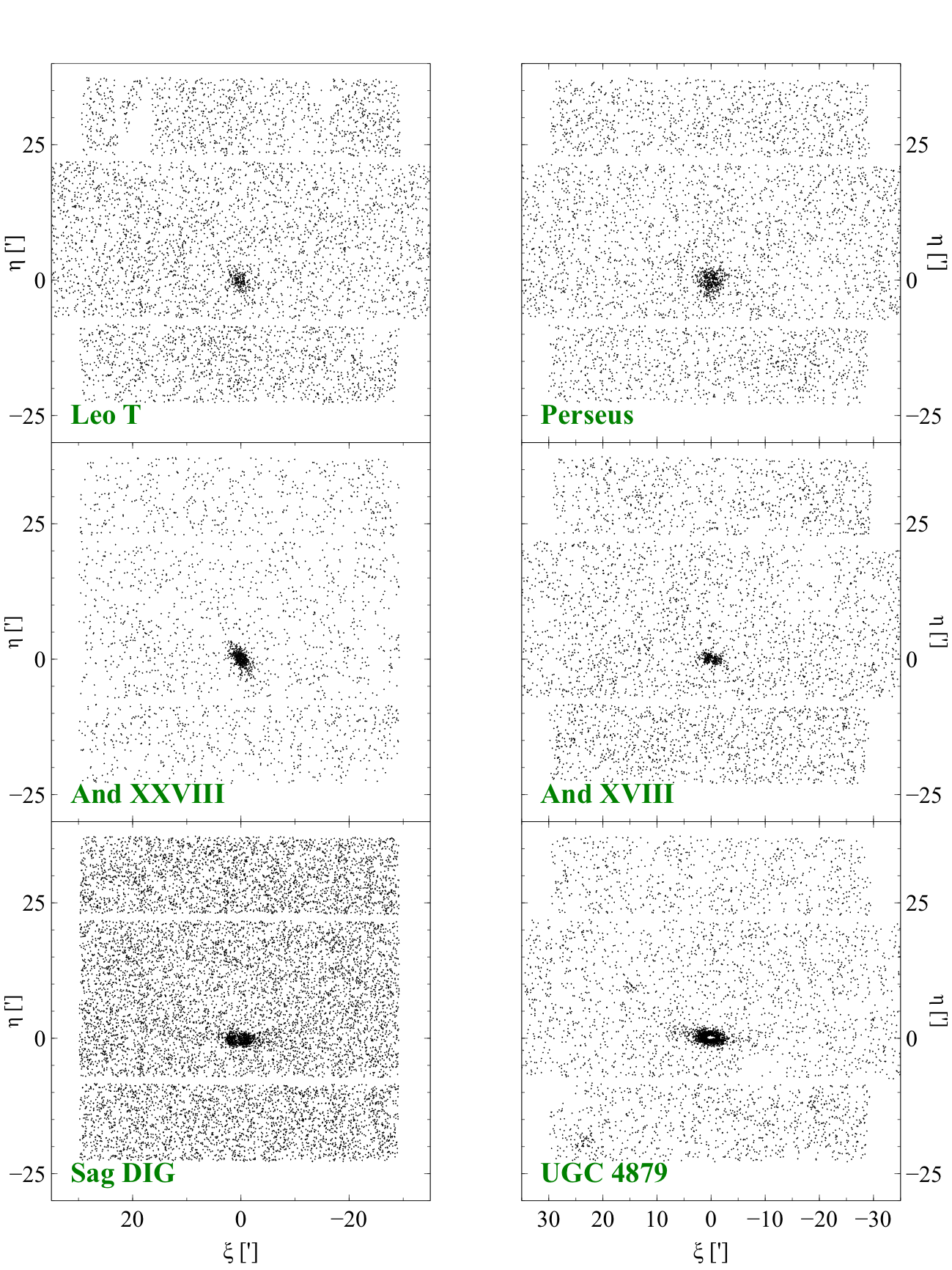}
    \caption{Spatial distribution of RGB stars for 6 of the dwarfs in the Local Group subset over the entire MegaCam field of view. The remaining dwarfs are shown in Figure \ref{fig:spatial2}.}
    \label{fig:spatial1}
\end{figure*}

\begin{figure*}
	\includegraphics[width=.75\linewidth]{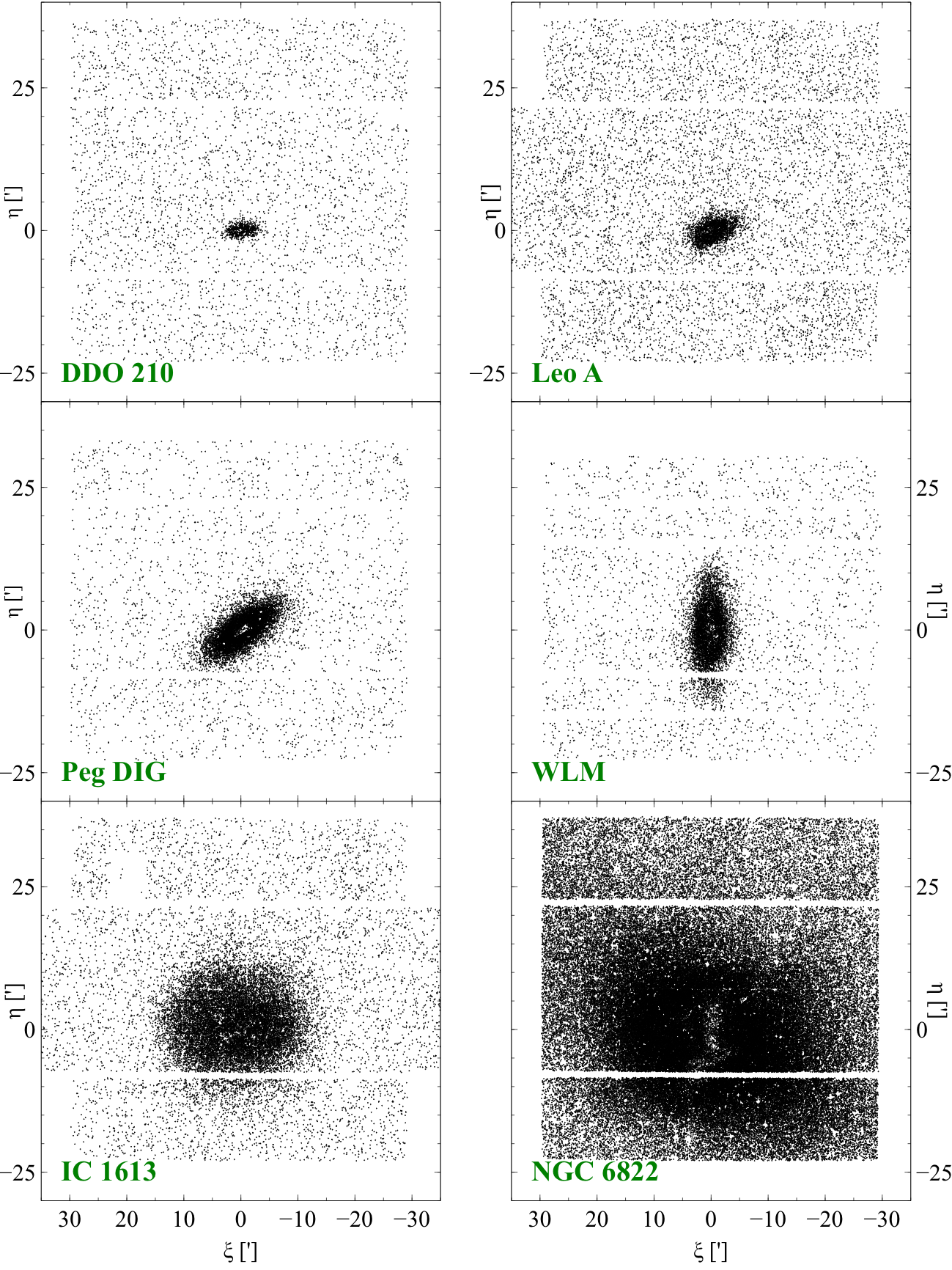}
    \caption{Spatial distribution of RGB stars for 6 of the dwarfs in the Local Group subset over the entire MegaCam field of view. The remaining dwarfs are shown in Figure \ref{fig:spatial1}.}
    \label{fig:spatial2}
\end{figure*}

Once the distance has been calculated for each dwarf galaxy, we examine their spatial distribution based on the RGB stars. These stars are selected  from the full stellar catalogue (no spatial constraints) as within the tram-lines previously identified and fainter than the TRGB. 
Figures \ref{fig:spatial1} and \ref{fig:spatial2} show the distribution of these stars in the entire field of view for each dwarf in the subset. The fields of view were positioned so that the galaxies were slightly off-center to avoid the small detector gaps between CCDs. The two large gaps in the detector between CCD rows 1 \& 2 and rows 2 \& 3 are clearly visible. Note that those observations taken with the newer filter set at CFHT have a different footprint. The new filters are physically larger, and the ``ears" of the detector can now be used whereas previously they were not (increasing the total number of science CCDs from 36 to 40). 

The main panels of Figures~12 and 13 show the corresponding density maps for each of our dwarfs, where contour levels are set at a minimum of $2\,\sigma$ above the background, and pixels are 1 arcmin in each dimension. To estimate the background, we examine the density of stars outside of an ellipse corresponding to $5$~x~$R_s$ centered on the galaxy (where $R_s$ is the S\'{e}rsic scale radius fitted to the radial profile as described in Section~\ref{sec:fit}). This criteria is modified for the largest three galaxies as $5$~x~$R_s$ does not lie with in $1^{\circ}$ field of view. For WLM and IC 1613, we estimate the background at 3 and 2~x~$R_s$ respectively. For NGC 6822, no background is estimated; RGB stars belonging to the dwarf dominate over the entire full field of view.

Considering each galaxy as an ellipsoidal distribution of stars, the appropriate position angle and ellipticity is calculated for each galaxy using the moments of the corresponding unsmoothed RGB density maps. Specifically, from \cite{McConnachieIrwin2006}, the position angle (P.A.) and ellipticity ($e$) are given by:

\begin{eqnarray} \label{momeq}
P.A.=\frac{1}{2}arctan\left( \frac{2\sigma_{xy}}{\sigma_{yy}-\sigma_{xx}}\right) \\
e=\frac{\sqrt{(\sigma_{xx}-\sigma_{yy})^2+\sigma_{xy}^2}}{\sigma_{yy}+\sigma_{xx}}
\end{eqnarray}
where $\sigma_{xx},\sigma_{xy}$, and $\sigma_{yy}$ are the intensity weighted second moments. These moments are given by:
\begin{eqnarray}
\sigma_{xx}=\frac{\sum_{i}(x_{i}-X)^2I_i}{I_{tot}}\\
\sigma_{yy}=\frac{\sum_{i}(y_{i}-Y)^2I_i}{I_{tot}}\\
\sigma_{xy}=\frac{\sum_{i}(x_{i}-X)(y_{i}-Y)I_i}{I_{tot}}
\end{eqnarray}
where $(X,Y)$ is the center of the galaxy, and $(x_i, y_i)$ is the location of the $i^{th}$ pixel with intensity $I_i$.

Starting with a circular aperture, the P.A., $e$, and center are determined as a function of radius, in 0.5' steps, using these relationships. The analysis is then repeated, but this time only using those pixels that are contained within an ellipse with the newly derived P.A. and $e$. This process is repeated until the values converge. 

The resulting position angle and ellipticity profiles are shown as a function of semi major axis radius (SMA) in the side panels of Figure~12 and 13. These profiles are fairly regularly behaved in the uncrowded regions of the galaxy (indicated by the unshaded regions in Figures 12 and 13) for all but the faintest galaxies. A single P.A and $e$ is adopted for each galaxy by determining the values between $R_s$ and 3 $ \times R_s$ for use the the following analysis. We include the standard deviation within this range as uncertainties on these quantities . 

Radial profiles based on star counts alone are determined using the median values for P.A., ellipticity and center. When computing the radial profile, a mask is used to ignore areas of our field with chip gaps, edge regions, and areas where detection of stars was not possible due to the presence of highly saturated foreground stars. 

In all cases except NGC 6822, the profile is background subtracted using the background values estimated previously, generally using the area beyond $5 R_s$. The appropriate multiplicative factor for $R_s$, however, is estimated in the next section. In practice, this resulted in an iterative process, whereby we adopted an initial value for $R_s$ to estimate the background, and then derived a new value of $R_s$ based on the resulting profile. The entire process was repeated and continued until convergence. The final radial profiles based on star counts alone are shown in Figure \ref{fig:rgbprof}. These radial profiles clearly extend to larger radii that the integrated light profiles shown in Figure \ref{fig:intprofs}. The impacts of crowding and the resulting incompleteness of the stellar catalogues is apparent at small radii, most strongly for the more massive dwarfs (for example, WLM very clearly shows this feature). In the case of NG 6822, the dwarf dominates over background sources across the entire image so no background is subtracted. The resulting $R_s$ is still robustly measured.

\begin{figure*}
	\includegraphics[width=\linewidth]{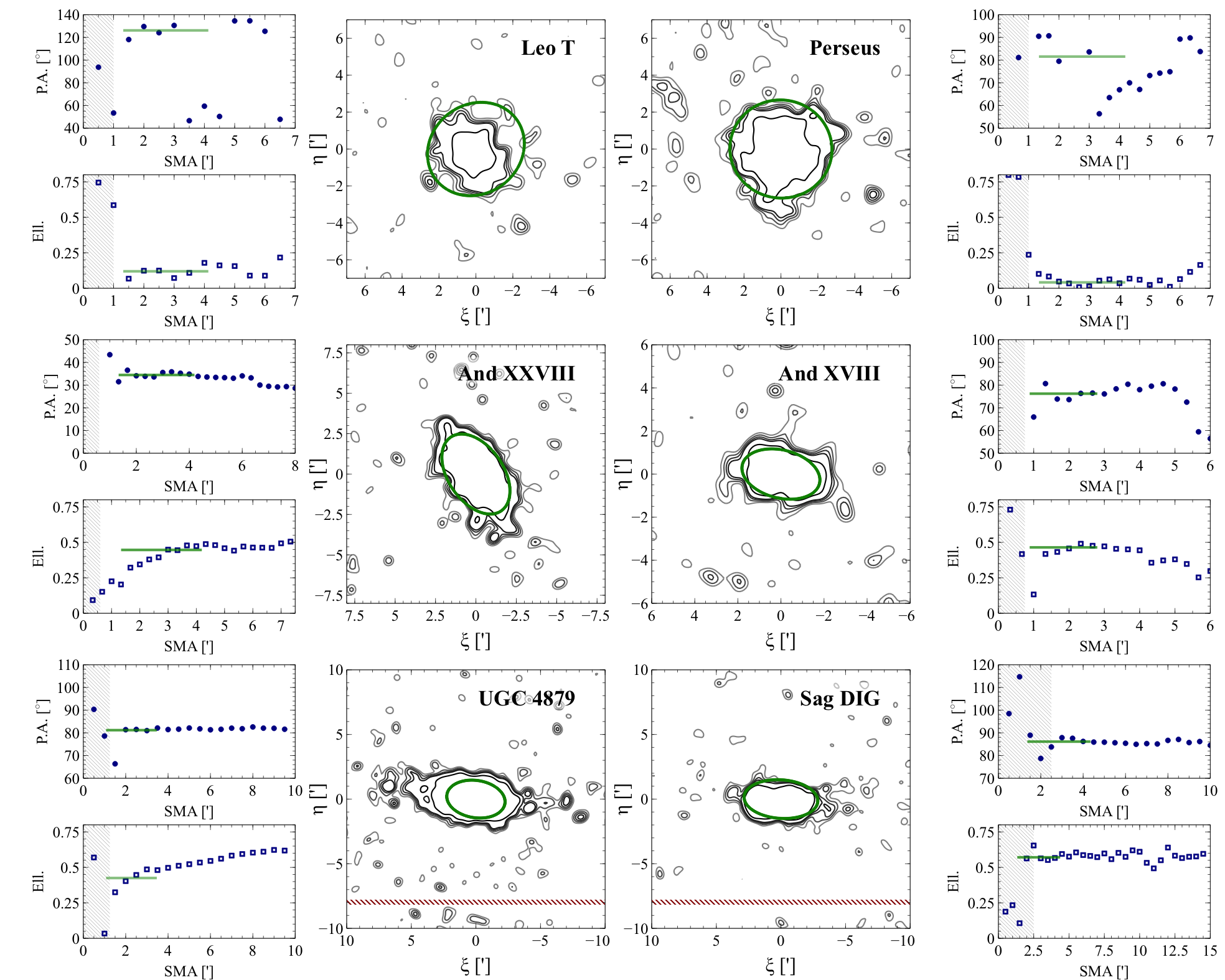}
    \caption{ For each dwarf, the upper and lower small panels show the position angle (P.A.) and ellipticity (Ell.) as a function of radius (Semi Major Axis - SMA). The green line indicates the adopted median from 1 to 3$R_s$.  The large panel shows a zoomed-in region showing RGB stars in contours. Contours of RGB stars are at 2,3,4,5,10,25 $\sigma$ about the estimated background. An ellipse (green) is shown with the median position angle and ellipticity with a semi major axis of $2R_s$. The grey shaded regions show the radii at which crowding is known to be significant, as shown in Figure \protect\ref{fig:rgbprof}. The red dash regions show the locations of the large chip gaps in the detector. }
    \label{fig:spat_1}
\end{figure*}

\begin{figure*}
	\includegraphics[width=\linewidth]{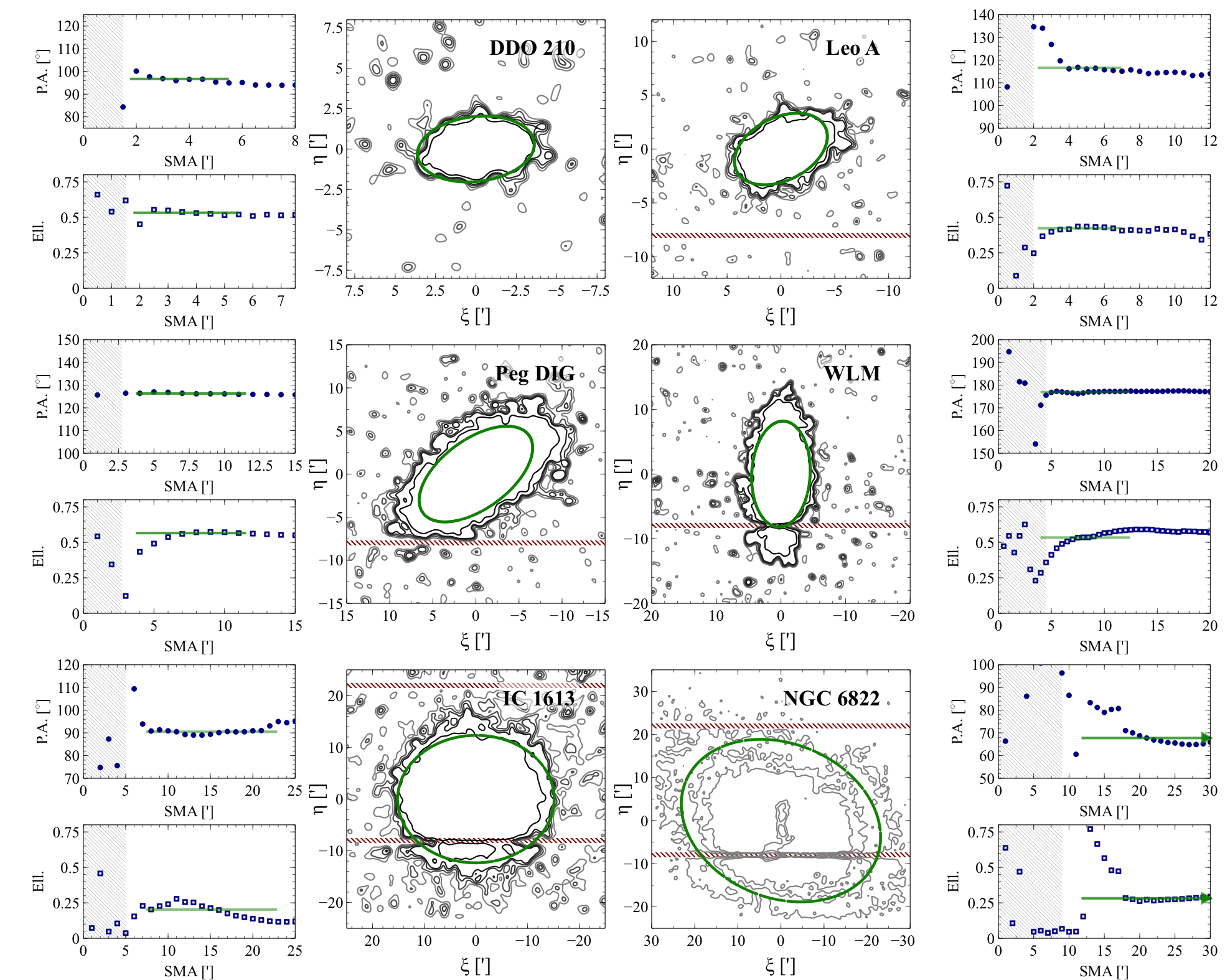}
    \caption{The same as Figure \ref{fig:spatial1}. For each dwarf, the upper and lower small panels show the position angle (P.A.) and ellipticity (Ell.) as a function of radius (Semi Major Axis - SMA). The green line indicates the adopted median from 1 to 3$R_s$.  The large panel shows a zoomed-in region showing RGB stars in contours. Contours of RGB stars are at 2,3,4,5,10,25 $\sigma$ about the estimated background. NGC 6822 is dominated across the full field by RGB stars so contours shown are at 10\%, 20\% and 30\% of the maximum stellar density. An ellipse (green) is shown with the median position angle and ellipticity with a semi major axis of $2R_s$. The grey shaded regions show the radii at which crowding is known to be significant, as shown in Figure \protect\ref{fig:rgbprof}.  The red dash regions show the locations of the large chip gaps in the detector.} 
    \label{fig:spat_2}
\end{figure*}

\begin{figure*}
	\includegraphics[width=0.9\linewidth]{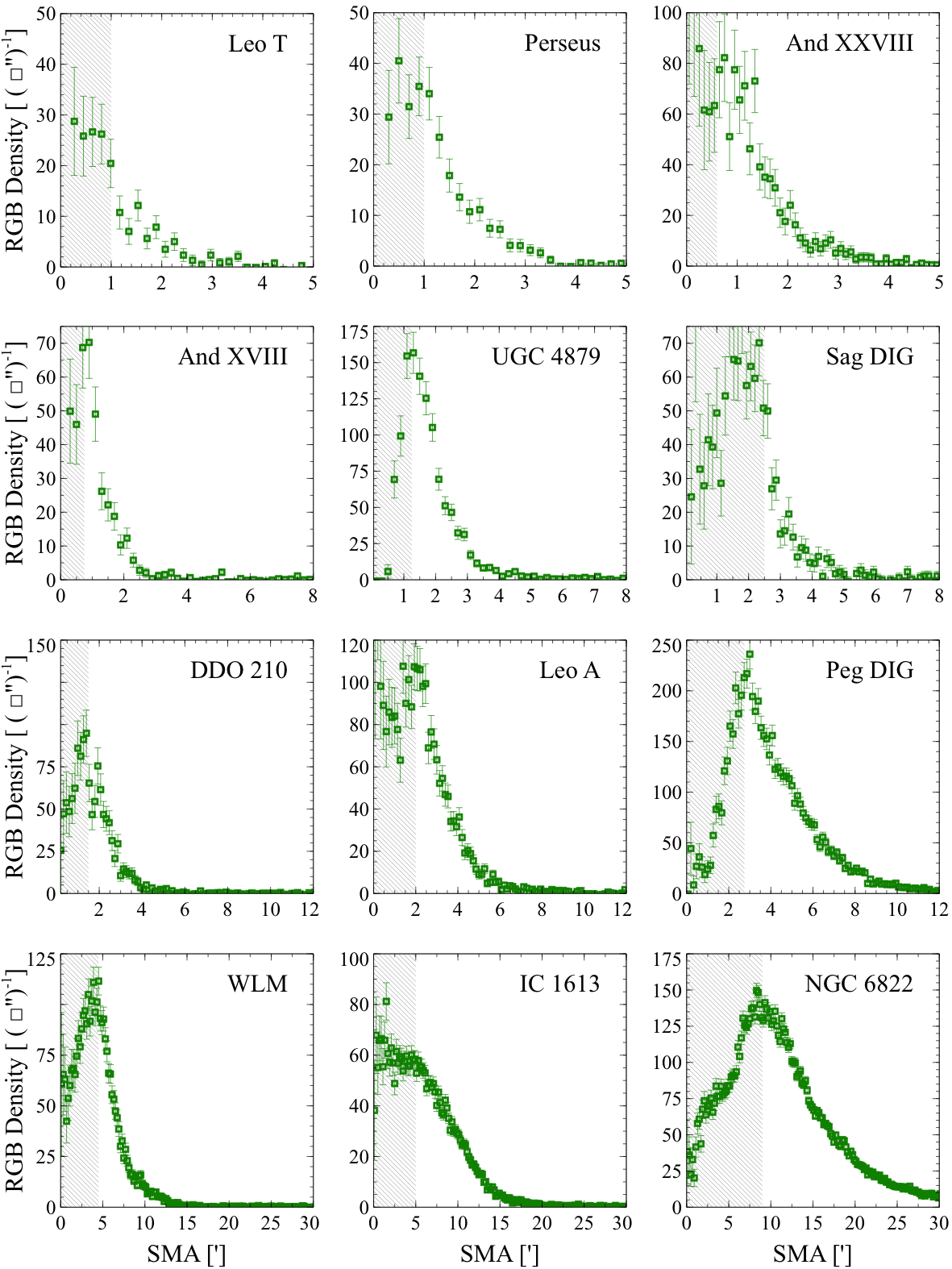}
    \caption{RGB stellar density profiles for each dwarf as a function of (SMA - semi major axis). The grey shaded region indicates where crowding is significant, and were excluded from our analysis.} 
    \label{fig:rgbprof}
\end{figure*}

\section{Dwarf galaxy profiles}\label{sec:fit}

\subsection{Profile fitting}

 Extended radial profiles are now generated using the RGB stellar density profiles as tracers of the outskirts of the galaxies and the integrated light profiles as tracers of the inner regions of the galaxy. A significant assumption behind this approach is that the integrated light and the RGB stars trace the same populations. This is clearly incorrect in detail. However, it is a reasonable and necessary assumption to connect the outer stellar profile to the inner integrated light region. The systematic effects are limited by matching the $i-$band profile to the RGB stars, since younger stars contribute more heavily to bluer bands. For some of the smaller dwarfs - And XXVIII, Leo T and Perseus - the dwarf does not appear in the $i-$ band via integrate light and we can only detect the integrated light in the $g-$ band. For these dwarfs, we instead match the stellar density profiles in the $g-$ band and adjust them to the $i-$band using the median colour of the RGB from the CMD for these dwarfs.  

The RGB stellar density profiles, measured in stars per unit area (derived in Section \ref{sec:stucture}) are matched with the integrated light profiles, measured in flux per unit area (as derived in Section \ref{sec:intlight}) by determining the appropriate scaling factor, $\gamma_{rgb}$. This factor is a multiplicative term that includes a factor due to the difference in units as well as containing information on the contribution of the RGB stars to the total integrated light of the galaxy. We fit $\gamma_{rgb}$ as an unknown parameter alongside the parameters for a S\'{e}rsic function (e.g. \citealt{Graham2005}):

\begin{equation}\label{S\'{e}rsic}
I(r)=I_e exp\left(-b_n\left[\left(\frac{r}{R_s}\right)^{\frac{1}{n}} -1\right] \right) \;
\text{where} \; b_n=1.9992n-0.3271
\end{equation}

\begin{figure*}
	\includegraphics[width=0.9\linewidth]{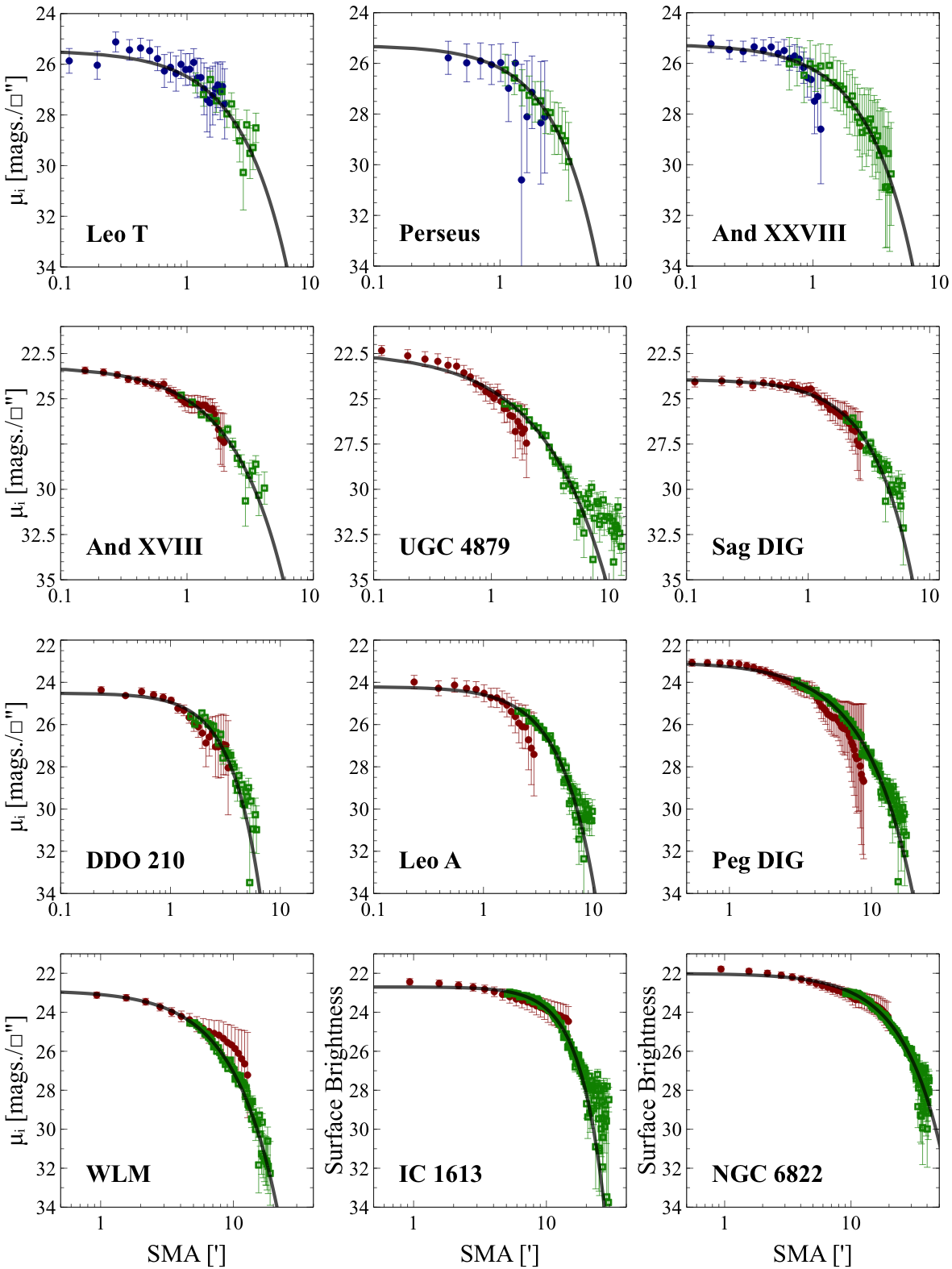}
    \caption{Combined radial profiles using integrated light (red circles) and star counts (green open squares) as a function of radius (SMA - semi major axis), for each galaxy. Overlaid are the best-fit S\'{e}rsic profiles (black line). For Leo T, Perseus, and And XXVIII no integrated light is detected in the $i-$ band so the profile is matched to the integrated $g$ profile and then corrected with the median colour (in the panels above, blue is adjusted $g-$ band, red is $i-$ band). } 
    \label{fig:intlight}
\end{figure*}

\noindent $I_e$ is the intensity at the effective (S\'{e}rsic) radius ($R_s$), and $n$ is the S\'{e}rsic index, which relates to the concentration of the galaxy. We use {\it emcee} \citep{ForemanMackey2013} to determine the most probable values of all these unknown parameters, and this allows us to explore any degeneracy between the structural parameters and the $\gamma_{rgb}$. The resulting radial profiles are shown in Figure \ref{fig:intlight} and overlaid are the best-fit S\'{e}rsic functions. 

\subsection{Two component models}
Each radial profile is also fit with the sum of two S\'{e}rsic profiles (again including $\gamma_{rgb}$ as a free parameter) to explore any evidence for more than one component in any of these galaxies. Both Bayesian and Akaike Information Criteria (BIC, AIC, respectively; \citealt{Kass1995}) are computed to determine whether the extra parameters in the multi-component model are statistically warranted. Specifically, we compute these parameters for both the single and double-component profiles, and look at the difference. Models with lower BIC and AIC values are formally preferred, and the specific difference between the two sets of values indicates to what degree the model is preferred. These two criteria differ in how goodness-of-fit is weighted verses the number of parameters. Broadly speaking, the AIC is likely to favour models which fit well even if there are more parameters, and the BIC is likely to favour simpler models that fit less well. 

\begin{figure}
	\includegraphics[width=0.9\linewidth]{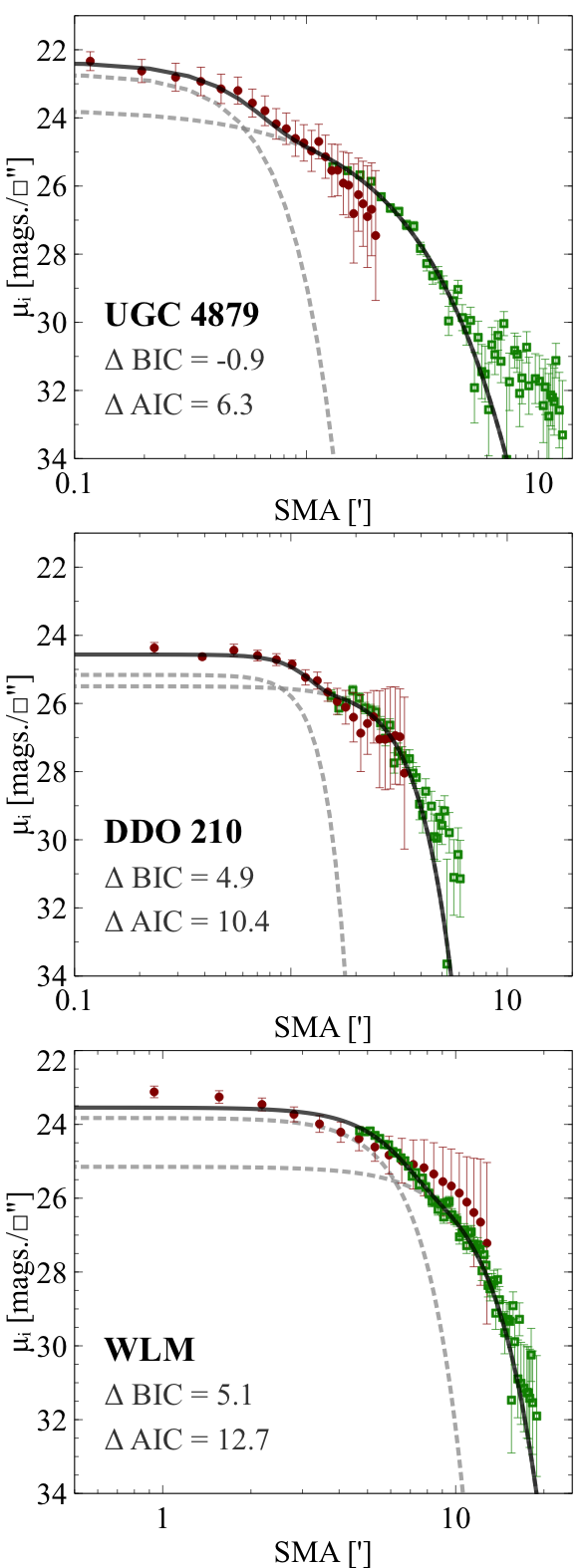}
    \caption{Combined radial profiles using integrated light (red circles) and star counts (green open squares) as a function of radius (SMA - semi major axis) for those galaxies for which there is some evidence of 2 components. Overlaid as solid lines are the best-fitting two component S\'{e}rsic profiles (black line), with the  individual components shown as dashed grey lines. Also indicated are the AIC and BIC for each galaxy.} 
    \label{fig:2comps}
\end{figure}

Comparison of the AIC and BIC values for each dwarf for the single- and double-component models indicate that most dwarfs are clearly more suitably described by single component models. UGC 4879, DDO 210 and WLM are exceptions for which the preferred model is less clear. Figure \ref{fig:2comps} shows the two-component S\'{e}rsic fits for each of these systems. Also shown are the values of $\Delta(BIC) = BIC_1 - BIC_2$ and $\Delta(AIC) = AIC_1 - AIC_2$ for each galaxy, where the subscript refers to the number of components fit.

For the BIC statistic, a difference in values less than around 2 does not imply any significant preference between models, whereas differences of order 2 - 5 are indicative of moderate preference for the model with the lower BIC value. For the AIC statistic, the exponential of half of the difference in values is broadly proportional to the probability that the model with the lower AIC value minimises the information loss. In our case, this implies that, for UGC 4879, the one component model is a factor of $\exp(-6.3/2) = 0.04$ more likely than the two component model to minimise the information loss (i.e., the two component model is significantly preferred); for DDO 210 and WLM, the corresponding numbers are $5.5 \times 10^{-3}$ and $1.7 \times 10^{-3}$.

Interpretation of these results requires some caution. In particular, stellar population gradients could have a very strong effect on these profiles, because of our necessary reliance on integrated light studies for the central regions. Inspection of the CMDs of these galaxies in Figures~8 and 9 shows that young stars are present in these galaxies, and appear especially strong in the magnitude range of our observations in WLM and DDO 210. Indeed, the differences in the radial gradients of young blue stars and RGB stars has been measured by \cite{McConnachie2006}, \cite{Leaman2012} and \cite{Jacobs2011} for DDO 210, WLM and UGC 4879, respectively. Thus, it is quite possible that the preference for two components in our analysis for these galaxies is in a result of these differences in the spatial distribution of the stellar populations, although we note that the very few bright blue stars in our CMD for UGC 4879 suggest other factors potentially at play. Indeed, it has previously been argued (\citealt{Bellazzini2011a}) that UGC 4879 may host a disk-like component because of the "wings" that are visible in the 2D surface brightness map (see Figure \ref{fig:spat_1}). We will discuss this feature in more detail in Section~6. 

\subsection{Derived structural parameters}

For each galaxy, Table \ref{tab:momtable} lists our derived positional and shape parameters, specifically: RA, Dec (from \citealt{McConnachie2012}), distance modulus, distance, position angle, ellipticity, as well as offsets of the center of the RGB stellar distribution from the adopted centers.

Table \ref{tab:Sersictable} lists the fitted S\'{e}rsic parameters for each dwarf, specifically $\mu_e$, $R_s$, $n$, as well as the adopted background density of RGB stars. Also included are the best fit parameters for the two component S\'{e}rsic profiles for DDO 210, UGC 4879 and WLM. 

Table \ref{tab:derivetable} shows derived parameters from the S\'{e}rsic fits following \cite{Graham2005}: $\mu_o$ is the central surface brightness, $\langle\mu\rangle_e$ is the mean surface brightness inside the effective radius, $m_{tot}$ is the apparent total magnitude, and $M_{tot}$ is the absolute magnitude using the distance modulus given in Table~\ref{tab:momtable}.

\begin{table*}
	\centering
	\caption{Derived positional and shape information for each dwarf galaxy. RA and Dec. for each galaxy are taken from \protect\cite{McConnachie2012}.}
	\label{tab:momtable}
	\begin{tabular}{lccccccccc} 
	\hline
	\hline
  Name      &   RA      &  Dec. & $(m-M)_o$ & Distance & $P.A.$ & $e$ & $\Delta x$ & $\Delta y$ \\
    &&&& [kpc] &[$^{\circ}$]  &  &   [']     & [']   \\
	\hline
	Leo T        & $09^{h}34^{m}53.4^{s}$ &  $+17^{o}03{'}05"$ &  23.08$\pm$0.08&        413$\pm$15&        121.13$\pm$34.66&        0.12$\pm$0.08&        -0.22$\pm$0.06&        0.20$\pm$0.04 \\ 
    Perseus      & $03^{h}01^{m}22.8^{s}$ &  $+40^{o}59{'}17"$ & 24.18$\pm$0.11&        686$\pm$34&        81.56$\pm$21.43&        0.04$\pm$0.08&        0.07$\pm$0.01&        0.18$\pm$0.02 \\ 
    And XXVIII   & $22^{h}32^{m}41.2^{s}$ &  $+31^{o}12{'}58"$ & 24.44$\pm$0.04&        772$\pm$16&        35.04$\pm$0.97&        0.42$\pm$0.06&        0.00$\pm$0.01&        -0.07$\pm$0.02 \\
    And XVIII    & $00^{h}02^{m}14.5^{s}$ &  $+45^{o}05{'}20"$ & 25.36$\pm$0.08&        1178$\pm$43&        75.12$\pm$4.47&        0.44$\pm$0.12&        -0.01$\pm$0.17&        -0.07$\pm$0.06 \\ 
    UGC 4879     & $09^{h}16^{m}02.2^{s}$ &  $+52^{o}50{'}24"$ & 25.42$\pm$0.06&        1213$\pm$33&        81.19$\pm$6.47&        0.43$\pm$0.06&        -0.23$\pm$0.02&        -0.07$\pm$0.02 \\
    Sag DIG      & $19^{h}29^{m}59.0^{s}$ &  $-17^{o}40{'}51"$ & 25.39$\pm$0.08&        1198$\pm$47&        86.91$\pm$3.44&        0.56$\pm$0.18&        -0.17$\pm$0.38&        0.35$\pm$0.05 \\
    DDO 210      & $20^{h}46^{m}51.8^{s}$ &  $-12^{o}50{'}53"$ & 24.97$\pm$0.09&        988$\pm$43&        96.58$\pm$1.44&        0.53$\pm$0.05&        0.25$\pm$0.09&        -0.04$\pm$0.01 \\
    Leo A        & $09^{h}59^{m}26.5^{s}$ &  $+30^{o}44{'}47"$ & 24.28$\pm$0.05&        717$\pm$17&        116.40$\pm$6.11&        0.42$\pm$0.05&        0.49$\pm$0.18&        0.12$\pm$0.10 \\ 
    Peg DIG      & $23^{h}28^{m}36.3^{s}$ &  $+14^{o}44{'}35"$ & 24.77$\pm$0.04&        898$\pm$19&        126.30$\pm$0.34&        0.56$\pm$0.05&        0.20$\pm$0.09&        0.12$\pm$0.04 \\
    WLM          & $00^{h}01^{m}58.2^{s}$ &  $-15^{o}27{'}39"$ & 24.85$\pm$0.05&        934$\pm$21&        177.02$\pm$0.50&        0.54$\pm$0.06&        -0.17$\pm$0.01&        0.05$\pm$0.10 \\ 
    IC 1613      & $01^{h}04^{m}47.8^{s}$ &  $+02^{o}07{'}04"$ & 24.18$\pm$0.06&        686$\pm$19&        90.49$\pm$1.02&        0.20$\pm$0.05&        -0.71$\pm$0.27&        -0.90$\pm$0.19 \\
	NGC 6822     & $19^{h}44^{m}56.6^{s}$ &  $-14^{o}47{'}21"$ & 23.78$\pm$0.05&        570$\pm$12&        66.88$\pm$14.93&        0.28$\pm$0.15&        -0.60$\pm$1.60&        1.08$\pm$0.80 \\
	\hline
	\hline

	\end{tabular}
\end{table*}

\begin{table*}
	\centering
	\caption{Best-fit S\'{e}rsic parameters for each dwarf, in addition to the adopted  background density of RGB stars.  Note that $\mu_e$ is  the equivalent of $I_e$ given in magnitudes. The two component fits (a sum of two S\'{e}rsic functions) are given for those dwarfs for which such a model appears plausible as discussed in the text. }
	\label{tab:Sersictable}
	\begin{tabular}{lccccccc} 
		\hline
		\hline
 		 Name & & $\mu_e$  &  $R_s$  & $n$  & C & \\
               && [mags/sq.arcsec] &['] && [RGB stars/sq.arcmin$^{-1}$] &\\
		\hline
	Leo T && 26.75 $\pm$ 0.26 & 1.39 $\pm$ 0.20 & 0.86 $\pm$ 0.36 & 1.43 $\pm$ 0.02 \\  
	Perseus && 26.96 $\pm$ 0.32 & 1.38 $\pm$ 0.19 & 0.79 $\pm$ 0.63 & 1.14 $\pm$ 0.02 \\  
    And XXVIII && 26.70 $\pm$ 0.36 & 1.38 $\pm$ 0.06 & 0.84 $\pm$ 0.13 & 0.64 $\pm$ 0.02 \\ 
    And XVIII && 24.91 $\pm$ 0.09  & 0.92 $\pm$ 0.05 & 0.95 $\pm$ 0.10 & 1.16 $\pm$ 0.02 \\ 
    UGC 4879$^{\star}$ && 24.77 $\pm$ 0.16 & 1.13 $\pm$ 0.10 & 1.28 $\pm$ 0.17 & 0.88 $\pm$ 0.02 \\  
    Sag DIG && 25.19 $\pm$ 0.09 & 1.43 $\pm$ 0.08 & 0.75 $\pm$ 0.07 & 3.39 $\pm$ 0.04 \\ 
    DDO 210$^{\star}$ && 25.48 $\pm$ 0.07 & 1.63 $\pm$ 0.06 & 0.61 $\pm$ 0.05 & 0.99 $\pm$ 0.02 \\ 
    Leo A && 25.42 $\pm$ 0.13  & 2.30 $\pm$ 0.09 & 0.72 $\pm$ 0.07 & 1.89 $\pm$ 0.03 \\  
    Peg DIG && 24.36 $\pm$ 0.05 & 3.81 $\pm$ 0.05 & 0.77 $\pm$ 0.03 & 0.64 $\pm$ 0.03 \\  
    WLM$^{\star}$ && 24.20 $\pm$ 0.07  & 4.10 $\pm$ 0.13 & 0.77 $\pm$ 0.04 & 0.72 $\pm$ 0.02 \\  
    IC 1613 && 23.29 $\pm$ 0.07 & 7.57 $\pm$ 0.05 & 0.43 $\pm$ 0.02 & 1.36 $\pm$ 0.03 \\  
    NGC 6822 && 23.25 $\pm$ 0.05 & 11.95 $\pm$ 0.07 & 0.73 $\pm$ 0.02 & --\\
\hline
UGC 4879 & Inner & 23.32 $\pm$ 0.33 & 0.36 $\pm$ 0.09 & 0.45 $\pm$ 0.20 & 0.88 $\pm$ 0.02  & \\
& Outer & 25.37 $\pm$ 0.25 & 1.36 $\pm$ 0.10 & 0.91 $\pm$ 0.15 &0.88 $\pm$ 0.02 & \\
DDO 210 & Inner & 25.26 $\pm$ 0.17 & 0.69 $\pm$ 0.08 & 0.22 $\pm$ 0.05 & 0.99 $\pm$ 0.02 \\
& Outer & 25.93 $\pm$ 0.15 & 1.68 $\pm$ 0.06 & 0.37 $\pm$ 0.05 & 0.99 $\pm$ 0.02 \\
WLM & Inner & 24.12 $\pm$ 0.12 & 3.66 $\pm$ 0.19 & 0.30 $\pm$ 0.06 & 0.72 $\pm$ 0.02 \\
& Outer & 25.56 $\pm$ 0.23 & 6.35 $\pm$ 0.38 & 0.36 $\pm$ 0.07 & 0.72 $\pm$ 0.02 \\
\hline
\hline
	\end{tabular}

\end{table*}

\begin{table*}
	\centering
	\caption{Derived parameters for each dwarf, following the relationships given in \protect\cite{Graham2005}, $\mu_{i,o}$ is the central surface brightness in the i-band,   $\langle\mu\rangle_{i,e}$ is the mean surface brightness inside the effective radius in the i-band, $m_{i, tot}$ is the total apparent magnitude in the i-band, $(g - i)$ is the median colour, and $M_{i, tot}$ is the absolute i-band magnitude. }
	\label{tab:derivetable}
	\begin{tabular}{lccccccc} 
		\hline
		\hline
 		 Name   & $\mu_{i,o}$  & $\langle\mu\rangle_{i,e}$  & $m_{i, tot}$ & $(g - i)$ & $M_{i, tot}$ \\
		\hline
Leo T&25.24 $\pm$ 0.82 & 26.14 $\pm$ 0.33 & 14.69 $\pm$ 0.34 & 0.61 $\pm$ 0.24 & -8.39 \\ 
Perseus&25.6 $\pm$ 1.4 & 26.39 $\pm$ 0.61 & 14.9 $\pm$ 0.62 & 1.13 $\pm$ 0.43 &-9.28 \\ 
And XXVIII&25.23 $\pm$ 0.46 & 26.08 $\pm$ 0.37 & 15.1 $\pm$ 0.38 & 0.68 $\pm$ 0.32 &-9.34 \\ 
And XVIII&23.2 $\pm$ 0.23 & 24.24 $\pm$ 0.10 & 14.17 $\pm$ 0.28 & 0.91 $\pm$ 0.12 &-11.19 \\ 
UGC 4879&22.35 $\pm$ 0.4 & 23.96 $\pm$ 0.17 & 13.42 $\pm$ 0.21 &0.97 $\pm$ 0.36 & -12.0 \\ 
Sag DIG&23.92 $\pm$ 0.18 & 24.62 $\pm$ 0.1 & 13.85 $\pm$ 0.56 & 0.52 $\pm$ 0.33 &-11.54 \\ 
DDO 210&24.51 $\pm$ 0.13 & 25.0 $\pm$ 0.08 & 13.87 $\pm$ 0.14 &0.60 $\pm$ 0.73 & -11.1 \\ 
Leo A&24.21 $\pm$ 0.20 & 24.87 $\pm$ 0.14 & 12.77 $\pm$ 0.17 & -0.15 $\pm$ 0.07 &-11.51 \\ 
Peg DIG&23.04 $\pm$ 0.08 & 23.78 $\pm$ 0.05 & 10.88 $\pm$ 0.14 &0.84 $\pm$ 0.28 & -13.89 \\ 
WLM&22.88 $\pm$ 0.11 & 23.62 $\pm$ 0.07 & 10.51 $\pm$ 0.17 & -0.16 $\pm$ 0.26 &-14.34 \\ 
IC 1613&22.71 $\pm$ 0.08 & 22.93 $\pm$ 0.07 & 7.9 $\pm$ 0.1 &1.47 $\pm$ 0.27 & -16.28 \\ 
NGC 6822&22.02 $\pm$ 0.07 & 22.69 $\pm$ 0.05 & 6.79 $\pm$ 0.26 &1.38 $\pm$ 0.26 & -16.99 \\ 

\hline
\hline
	\end{tabular}
\end{table*}

\section{Discussion}\label{sec:results}

\subsection{Structures and substructures}

Together, the 2D and 1D surface brightness profiles presented in Figures~8, 9, 12 and 13 provide considerable information on the extended structures of all these isolated galaxies. Figures 12 and 13 show that nearly all of the galaxies have a relatively regular appearance, essentially appearing as large spheroids with no evidence of large scale asymmetries like tidal tails. For almost all of the dwarfs, the best-fitting S\'{e}rsic profile to the RGB distribution yields $n\sim 0.7$, i.e., the stellar distribution falls off more slowly than an exponential. This is fairly typical of many dwarf galaxies (e.g. \citealt{Ferrarese2006, Gerbrandt2015,Munoz2018b}). 

An exception to both of the above generalisations is UGC 4879. Here, the stellar distribution falls off faster than an exponential ($n \simeq 1.3$), and the galaxy does present evidence of extended "wings" along its major axis. This latter feature has previously been interpreted as an edge-on disk. The tentative evidence of two components in Figure~16 also suggests that a multi-component interpretation could be correct for this galaxy, where both components are dominated by old stars (potentially in contrast to DDO 210 and WLM; see discussion for WLM below). The extreme isolation of this system makes it extremely unlikely that such signatures are the results of a tidal interaction with a more massive galaxy (M31 or the MW). However, it is possible that this type of structure - especially the surface brightness map - could also be the result of a merger in this galaxy. Numerous observational studies in recent years (e.g., \citealt{MartinezDelgado2012, Paudel2017, Zhang2020}) - including in the Local Group (e.g., \citealt{Nidever2013, Amorisco2014})- suggest that dwarf-dwarf mergers are not uncommon. In addition to UGC 4879, Sag DIG also has some evidence of a weak extension in its outer isophotes on the eastern end of its major axis (also discussed in Paper~I).

Beyond the more distinct ``wings" of UGC 4879 and perhaps Sag DIG, the outskirts of several dwarfs are not ideally characterized by the Sersic fit (e.g. DDO 210, Leo A, IC 1613). In theses cases, an excess of stars is found in the outskirts, rather than a deficit. \citealt{Irwin1995} discuss these ``extra-tidal" stars extensively in the context of Milky Way dwarf spheroidals, whether these stars are being stripped or are evidence that the system is not in dynamic equilibrium. For {\it Solo} dwarfs, the likelihood of tidal interactions is reduced (though not eliminated) by their isolation. Interestingly, the excess of stars found in UGC 4879 at large radius persists in both the single and two component fits.

\cite{Wheeler2015} suggest that dwarf galaxies such as the {\it Solo} dwarfs could well host satellites within around 50\,kpc of the main dwarfs. \cite{Dooley2017} take this further and explicitly calculate the probability of there being a satellite found around each of the {\it Solo} dwarfs within the approximate observational footprint for each of our fields (see their Table~ 3). For reference, our MegaCam field reaches out to about 0.5 degrees from the centers of each galaxy; for the closest galaxy (Leo T) this corresponds to a maximum (projected) radius of around $3.5$\,kpc; for the most distant galaxy (UGC 4879), this corresponds to around $10$\,kpc. \cite{Dooley2017} predict that And XVII, And XXVIII, Sag DIG, Peg DIG and Leo A all have around a 10\% chance of having a satellite in this region, and WLM, IC 1613 and UGC 4879 all have around a 10 -- 20\% chance of having a satellite in this region. Notably, NGC 6822, as one of the most massive dwarfs, has a relatively low probability due to the relatively small radius out to which we probe it. 

No evidence of satellites or other concentrated substructures are found around any of the galaxies to our surface brightness limits (around 30 -- 32 mags\,sq.arcsec), as inspection of Figures 12 and 13 makes clear. Candidate features (e.g., the feature in the isophotes around Perseus at $(\xi \simeq -5', \eta \simeq -2.5')$ were examined and found to be consistent with fluctuations (statistical or due to irregular features e.g., holes due to bright foreground stars) in the background. Further, the OPTICS clustering algorithm (\citealt{Ankerst1999}; see its application to a similar dataset in \citealt{McConnachie2018}) was also applied to the RGB dataset, and no additional significant features were found. We note that this is relatively unsurprising, given that the probabilities given by \cite{Dooley2017} are not large for any single object. Further, we note that these probabilities correspond to satellites with stellar masses as low as $10^4\,M_\odot$. At this limit, detection would be very hard given the data in hand; the lowest stellar mass satellites  identified in PAndAS (which used the same instrument, same bands and broadly similar exposure times) were of order $ 10^{4.2}\,M_\odot$, although most were $>10^{4.5}\,M_\odot$ (see discussion in \citealt{Martin2016}). Examination of the numbers in Table~3 of \citealt{Dooley2017} suggests that the probability of discovering new satellites around some of the more distant, more massive galaxies in the rest of the {\it Solo} dataset remains good.

\subsection{Specific case studies}

Here we present a closer look at our results for three galaxies included in our sample - Sag DIG, WLM and DDO 210 - and use these as case studies to highlight aspects of our approach, interpretation of results, and comparison to previous studies. In particular, Sag DIG was already analysed in Paper I; WLM is a large and very well studied galaxy; DDO 210 has previously been studied with similar observations and methods and is at the faint end of our sample.  

\subsubsection{Sag DIG}

\begin{figure}
	\includegraphics[width=0.9\linewidth]{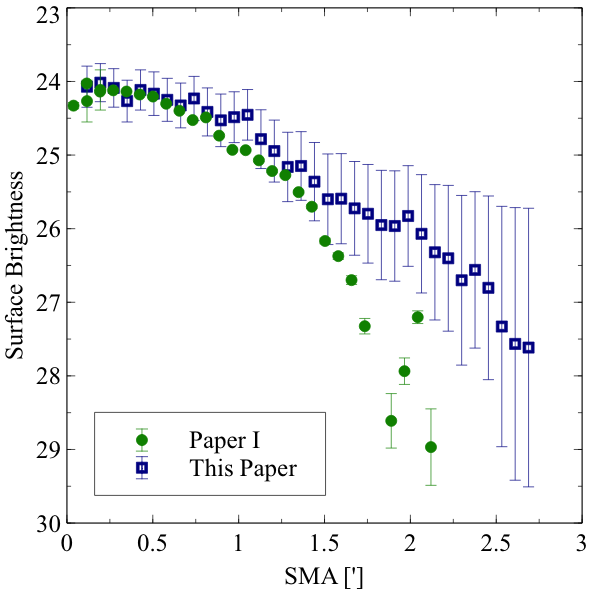}
    \caption{Comparison of the (background-subtracted) $i-$band integrated light profiles as a function of radius (SMA - semi major axis) for Sag DIG, derived in this paper(blue open squares) and in Paper~I (green circles). More robust masking of foreground sources in the current work leads to a more extended surface brightness profile. See text for details.} 
    \label{compareprofs}
\end{figure}

While Sag DIG was extensively studied in Paper I, there are a few important differences between the methodology applied in Paper I and in this current work. The integrated light and resolved RGB profiles were matched in Paper I by matching them "by eye" in the overlapping regions. Archival HST observations were used in order in straddle the region between the resolved stars and integrated light from the CFHT data. However, appropriate HST observations are not available for all targets, and so this method was not used in this current paper. Instead, we make the scaling parameter a free parameter and employ $emcee$ to obtain the best fits and explore correlations between parameters. 

Critically, the current work takes a more rigorous approach to dealing with the background in the integrated light analysis of the dwarf. In Paper I, masking was applied to all pixels that fell above a certain threshold of counts. In the current work, a threshold is still applied, but also pixels in the neighbourhood of these "hot" pixels are also masked, in order to suppress the effect of halos around bright stars and similar effects. This change in strategy has a significant effect for a galaxy like Sag DIG that is situated at relatively low Galactic latitude, and hence is impacted heavily by (bright) Milky Way foreground stars. In the CMD in Figure~8, this foreground is prominent, and even Galactic substructure due to the Sagittarius Stream can be seen.

Figure~\ref{compareprofs} shows the difference in the resulting integrated light profiles for Sag DIG. In particular, the more complete masking of bright sources in the current paper means that the overall background that is estimated for Sag DIG is smaller (by around 20~ADU, or just less than 1~\% of the previous background value). This changes the shape of the outer profile considerably, such that the new analysis clearly favors a more extended surface brightness profile. Note also that the uncertainties are much larger in this new analysis, since we explicitly take into account the systematic variation in the background due to the irregular distribution of foreground sources, whereas this was not the case in Paper~I. The net flow-down effect of this change in our analysis is considerable for Sag DIG ($R_s = 0.95 \pm 0.01$ in Paper I compared to $R_s = 1.43 \pm 0.08$\,arcmins in the current analysis). This emphasises the challenge of integrated light studies of these faint systems, and the need for a homogeneous approach to the analysis.

\cite{Beccari2014} also present a wide and deep study of the resolved stellar and HI structures of Sag DIG.  The S\'{e}rsic radius derived from \cite{Beccari2014} is approximately 1.1'; no error bars are given on this measurement but it appears to be in reasonable agreement to that derived here.  The TRGB distance estimate $(m-M)_o =25.39 \pm 0.08$ agrees well with $(m-M)_o =25.16 \pm 0.11$ from \cite{Beccari2014}. \cite{Beccari2014} and \cite{Lee2000} estimate the position angle to be $90^{\circ}$, which is in agreement with the value (measured East from North) found  in this work ($87^{\circ} \pm 3^{\circ}$). Similarly the ellipticity from \cite{Beccari2014} aand \cite{Lee2000} is $e=0.5$, which agrees with $0.56 \pm 0.18$ estimated here.

\subsubsection{WLM}
WLM (also known as DDO 221) is one of the more massive dwarfs in the {\it Solo} Survey. For comparison of the structural parameters, we focus on the detailed analysis from \cite{Leaman2012,Leaman2013}. This deep, wide field analysis also uses resolved RGB stars to study the older stellar populations in addition to including spectroscopy of 180 RGB stars. \cite{Leaman2012} derive the ellipticity and position angle of this dwarf by fitting ellipses using MPFITELLIPSE \cite{Markwardt2009}, finding $e=0.55$ and $P.A.=179 ^{\circ}$. These values are in good agreement with $e=0.54 \pm 0.06$ and $P.A.=177^{\circ} \pm 0.5^{\circ}$ derived here. The ellipticity and position angle of WLM are well defined and do not vary significantly with radius. 

The half light radius given for an exponential fit to the overall population in \cite{Leaman2012} is approximately $5.8 \pm 0.2'$, whereas the RGB-only profile has a corresponding radius of approximately $4.7'$; here, we find a slightly smaller half-light radius ($4.10'\pm0.13'$). As discussed earlier, there is some evidence that the overall profile of WLM is best described by a two-component S\'{e}rsic model. We urge caution here, since it is unclear if this is due to the significant number of young stars in WLM, which certainly have a different spatial distribution to the RGB stars used to create the resolved star profile (see \citealt{Leaman2012}). 

\subsubsection{DDO 210}

DDO 210 (Aquarius) is another well studied dwarf and, prior to the discovery of Leo T, was the smallest known galaxy with a significant gaseous component. We measure the TRGB at $(m-M)_o=24.97 \pm 0.09$, in excellent agreement with \cite{Cole2014} who measured $(m-M)_o=24.95 \pm 0.10$ and comparable to \cite{McConnachie2006} with $(m-M)_o=25.15 \pm 0.08$. The position angle and ellipticity are derived using the same method in \cite{McConnachie2006} and agree closely. \cite{McConnachie2006} found the ellipticity changed from 0.6 in the inner regions to 0.4 in the outskirts;  again, this agrees closely with the current work where we find a median ellipticity of $e=0.53\pm0.05$ and a very modest radial gradient. 

\cite{McConnachie2006} demonstrate the complex structure of this galaxy in terms of the differences in the radial gradients of its different stellar populations, and the apparent disconnect between the spatial properties of the stars and HI distribution. \cite{HermosaMunoz2020} obtained spectroscopic measurements for 53 RGB stars and determine the axis of rotation for the stellar component to be $139^{\circ} {^{+17}_{-27}} ^{\circ}$ in contrast with the HI rotation measured by \cite{Iorio2017} ($77.3^{\circ} \pm 15.2^{\circ}$). In our analysis, we find that this galaxy is possibly better described with a two component profile, but again it is hard to determine if this represents two "old" components, or whether it reflects the age segregation in the stellar populations noted by \cite{McConnachie2006}. 

\subsection{Comparing the suite of parameters to the literature}

\begin{figure*}
	\includegraphics[width=\linewidth]{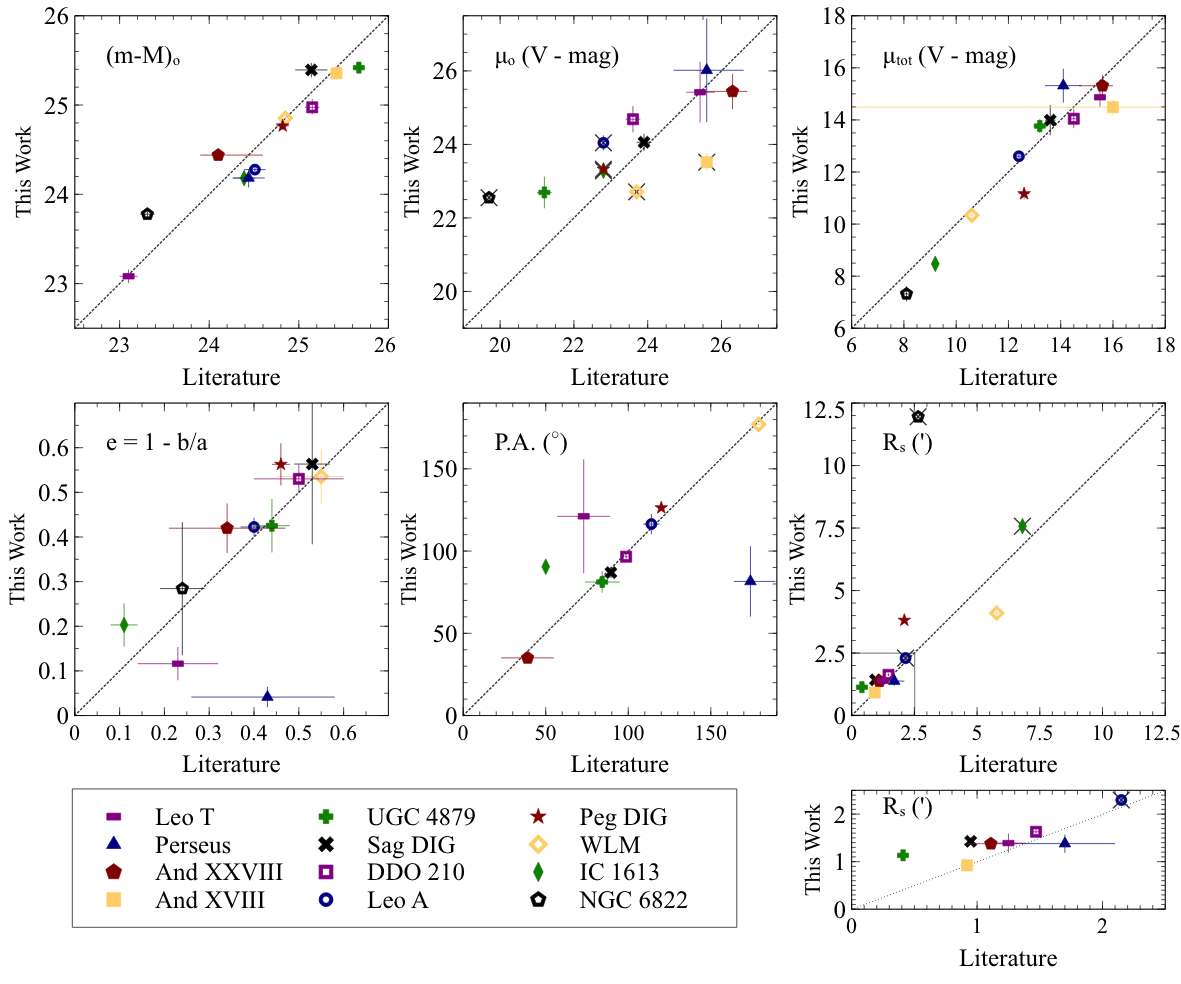}
    \caption{Comparison of various structural parameters derived in this paper compared to literature values, taken from the updated compilation in \protect\cite{McConnachie2012}. The small panel shows a zoomed version of the $R_s$ panel focused on small radii as indicated with the box. Points with a grey ``X" do not have robust literature error estimates. (And XXVIII: \citealt{Slater2011}, IC 1613: \citealt{Bernard2010,Devaucouleurs1991}, NGC 6822: \citealt{Gieren2006,Devaucouleurs1991,Dale2007}, Peg DIG: \citealt{McConnachie2005,Devaucouleurs1991}, Leo T:\citealt{Irwin2007,DeJong2008,Munoz2018b},WLM:\citealt{McConnachie2005,Devaucouleurs1991,Leaman2009}, And XVIII:\citealt{McConnachie2008,Conn2012}, Leo A:\citealt{Dolphin2002, Brown2007, Devaucouleurs1991, Vansevicius2004,Cole2007}, DDO 210:\citealt{McConnachie2005,McConnachie2006}, Sag DIG:\citealt{Higgs2016}, UGC 4879:\citealt{Jacobs2011,Kopylov2008,Bellazzini2011a}).}
    \label{fig:ellcomp}
\end{figure*}

We now compare the global set of parameters derived in this work to previously derived values, particularly the updated compilation of parameters from \cite{McConnachie2012}. Figure \ref{fig:ellcomp} shows the comparison for distance modulus ($(m-M)_o$), central surface brightness ($\mu_o$), apparent magnitude ($\mu_{tot}$), ellipticity ($e$), position angle (P.A.), and half-light radius($R_s$). For the central surface brightness and apparent magnitude, $\mu_o$ and $m_{tot}$ values in Table~\ref{tab:derivetable} measured in the $i-$ band were converted to their corresponding $g-$band values given the median $(g - i)$ values given in the same table. These values were then transformed to $V$ band using $V = g - 0.098 - 0.544(g - i)$ (Thomas et al., {\it submitted}). Points with an ``X" have literature values that are highly uncertain, and do not have quoted uncertainties. And XVIII does not have any previous value for its ellipticity or position angle (previous parameters were based on an image where the bulk of the object fell on a chip-gap).

It is worth emphasising that, to the best of our knowledge, some of these galaxies - especially NGC 6822, IC 1613 and Leo A - have not had detailed structural analyses conducted based on resolved stars in the modern era of CCD photometry. The integrated magnitudes of several of these galaxies date back to {\it The Third Reference Catalogue of Bright Galaxies} (\citealt{Devaucouleurs1991}), and in fact the half-light radii quoted in \cite{McConnachie2012} for these three galaxies are estimated by scaling other scale-radii that are measured by \cite{Devaucouleurs1991}. It is clear from the lower right panel of Figure~18 that this scaling is distinctly inconsistent for NGC 6822. 

Other notably deviant measurements include the central surface brightness of And XVIII, but this is relatively easily understood given that the earlier measurement was affected by an inconvenient chip-gap. Much more difficult to understand is the difference in ellipticity between our measurement of Perseus, which places the galaxy as nearly circular, compared to the measurement by \cite{Martin2013b}, which records a notable ellipticity. Inspection of the raw star count map in Figure~11 does not suggest that the galaxy is notably elliptical, consistent with our findings, and we speculate that the higher quality photometry for the MegaCam data compared to the discovery PS1 photometry has allowed a cleaner sample of member stars to be selected (this galaxy is at $b  \simeq -15$ degrees). Obviously, the poor agreement with the ellipticity measurement correlates with the poor agreement with the position angle measurement.

Generally, we see reasonable agreement between the bulk of our measurements and those from the literature. The only systematic difference that is obvious is that almost all of the effective radii that we derive are systematically larger than previous estimates. We suggest there are two primary factors that contribute. Firstly, some previous work has used observations with a smaller field of view. There is a non-trivial trend in the literature such that when viewing a galaxy out to larger radius, the galaxy is found to be larger than previously estimated. Secondly, and likely more important for the specific systems under study, there is a difference between the half-light radius of a galaxy, and a half-mass radius for any given tracer. Most {\it Solo} dwarfs are star forming, and so the central regions (and the light) are dominated by the younger, bluer brighter stars. However, our structural parameters are based on red giant branch stars, belonging to older stellar populations. It is often the case that older stellar populations are more extended than younger populations, and we stress that our results should be viewed as estimates of the older, halo-like populations of these dwarfs. The relative importance of these two factors (field of view verses half light/half mass) likely varies between dwarfs and the estimates used for comparisons. For example, Peg DIG and IC 1613 have a young stellar component \citep{Weisz2014} and $R_s$ in Figure \ref{fig:ellcomp} is from \cite{Devaucouleurs1991} which is based on the integrated light. As a result, the difference between the mass weighted and light weighted profiles is likely the dominant factor. This could also be a factor for some of the central surface brightness measurements (where integrated light studies will measure the flux density of all the stellar populations in the central regions, whereas our estimates are weighted towards the flux density due to older stellar populations). In contrast, structural parameters for And XVIII were previously based on observations in which the galaxy partially lies over a large chip gap in the MegaCam detector. Here, the differences seen in Figure \ref{fig:ellcomp} are likely largely due to the wider, more complete {\it Solo} observations. In addition, And XVIII has a predominately older SFH (as seen in \citealt{Weisz2014}), meaning the discrepancy between the mass and lighted weighted profiles is reduced. 

\section{Conclusion}
\label{sec:conclusions}

The purpose of this paper has been to determine a homogeneous set of structural parameters of isolated Local Group galaxies that is suitable for comparison with the well-studied Milky Way and M31 satellite populations. The diversity of this population of dwarf galaxies leads to considerable complexity in generating a homogeneous data set, but in general we find good agreement with previous studies. We have focused on their extended stellar structure and morphology and combined information from integrated light analysis of their inner regions with the resolved stellar populations of their outskirts, to parameterize their global structure to very faint surface brightness limits. Both the resolved stars and the integrated light require careful processing due the intrinsically small and faint nature of these dwarfs. Careful consideration of background subtraction is necessary due to the significant impacts on the resulting structure, particularly on the extended structure. The differences found in the structure of Sag DIG between Paper I and this work highlight these effects. 
All 12 galaxies are reasonably well described by 1D S\'{e}rsic functions, and no prominent stellar substructures, that could be signs of either faint satellites or recent mergers, are identified in the outer regions of any of systems examined.

Future contributions will use these results in comparison to surveys of Local Group satellite populations to examine differences between the populations of nearby dwarfs that may be attributed to the role of environment.

\section{Data Availability}

The raw data on which this analysis is based are publicly available in the CFHT archive (accessed via the Canadian Astronomical Data Center at \url{http://www.cadc-ccda.hia-iha.nrc-cnrc.gc.ca/en/cfht/}).

Processed data may be shared on reasonable request to the corresponding author.
 
\section{Acknowledgments}
Based on observations obtained with MegaPrime/MegaCam, a joint project of CFHT and CEA/DAPNIA, at the Canada-France-Hawaii Telescope (CFHT) which is operated by the National Research Council (NRC) of Canada, the Institut National des Sciences de l'Univers of the Centre National de la Recherche Scientifique (CNRS) of France, and the University of Hawaii. The observations at the CFHT were performed with care and respect from the summit of Maunakea which is a significant cultural and historic site. We respectfully acknowledge that UVic where this work was conducted is on the ancestral territories of Indigenous Peoples. As we explore the shared sky, we acknowledge our responsibilities to honour their past and enduring relationships with these lands. We thank Todd Burdulis for fantastic support throughout the {\it Solo} observational campaign.
CH kindly thanks the IoA for its support and hospitality while visiting during the completion of this work.
GB acknowledges financial support through the grant  (AEI/FEDER, UE) AYA2017-89076-P, as well as by the Ministerio de Ciencia, Innovación y Universidades (MCIU), through the State Budget and by the Consejería de Economía, Industria, Comercio y Conocimiento of the Canary Islands Autonomous Community, through the Regional Budget.


\bibliographystyle{mnras}
\bibliography{bib_ALL}



\bsp	
\label{lastpage}
\end{document}